\documentclass[12pt, a4paper]{article}

\baselineskip=\normalbaselineskip

\usepackage{mathrsfs,amsbsy,amssymb,latexsym,amsfonts,amsmath,amsthm}
\usepackage{graphicx,color}

\usepackage{bm}

\newcommand{\bx}{{\textbf{x}}}
\newcommand{\by}{{\textbf{y}}}

\newcommand{\bbC}{{\mathbb{C}}}
\newcommand{\bbR}{{\mathbb{R}}}

\makeatletter
\catcode`\@=11
\@addtoreset{equation}{section}

\def\@seccntformat#1{\csname the#1\endcsname.~~}
\makeatother

\begin{document}
\begin{titlepage}
\renewcommand{\thefootnote}{\fnsymbol{footnote}}
\begin{flushright}
KUNS-2783
\end{flushright}
\vspace*{1.0cm}

\begin{center}
{\Large \bf
Implementation of the HMC algorithm 
on the tempered Lefschetz thimble method
}
\vspace{1.0cm}

\centerline{
{Masafumi Fukuma${}^1$}%
\footnote{E-mail address: 
fukuma@gauge.scphys.kyoto-u.ac.jp},  
{Nobuyuki Matsumoto${}^1$}%
\footnote{E-mail address: 
nobu.m@gauge.scphys.kyoto-u.ac.jp} and
{Naoya Umeda${}^2$}%
\footnote{E-mail address: 
naoya.umeda1134@gmail.com}%
}

\vskip 0.8cm
${}^1${\it Department of Physics, Kyoto University, Kyoto 606-8502, Japan}\\
\vskip 0.1cm
${}^2${\it PricewaterhouseCoopers Aarata LLC, \\
Otemachi Park Building, 1-1-1 Otemachi, Chiyoda-ku, Tokyo 100-0004, Japan}
\vskip 1.2cm 

\end{center}

\begin{abstract}

The tempered Lefschetz thimble method (TLTM) is 
a parallel-tempering algorithm 
towards solving the numerical sign problem, 
where the system is tempered by the antiholomorphic gradient flow 
to tame both the sign and ergodicity problems simultaneously. 
In this paper, we implement the hybrid Monte Carlo (HMC) algorithm 
for transitions on each flowed surface, 
expecting that this implementation on TLTM will give a useful framework 
for future computations of large-scale systems including fermions. 
Although the use of HMC in Lefschetz thimble methods 
has been proposed so far, 
our crucial achievement here is 
that HMC is implemented on TLTM 
so as to work within the parallel-tempering algorithm in TLTM, 
especially by developing an algorithm 
to handle zeros of fermion determinants 
in the course of the molecular-dynamics process. 
We confirm that the algorithm works correctly 
by applying it to the sign problem of the Hubbard model on a small lattice, 
for which the TLTM is known to work with the Metropolis algorithm. 
We show that the use of HMC significantly reduces 
the autocorrelation times with less computational times 
compared to the Metropolis algorithm.

\end{abstract}
\end{titlepage}

\pagestyle{empty}
\pagestyle{plain}

\tableofcontents
\setcounter{footnote}{0}

\section{Introduction}
\label{sec:introduction}

The Markov chain Monte Carlo (MCMC) method has been an important tool 
in theoretical physics 
as it enables nonperturbative calculations of physical quantities. 
However, its application to some important research areas  
is still hindered due to the numerical sign problem. 
Examples include finite density QCD \cite{Aarts:2015tyj}, 
the quantum Monte Carlo simulations of 
strongly correlated electron systems \cite{Hirsch:1985,Loh:1990}, 
and real-time quantum field theories. 
Among many approaches towards solving the sign problem, 
algorithms using Lefschetz thimbles
\cite{Cristoforetti:2012su,Cristoforetti:2013wha, 
  Mukherjee:2013aga,Fujii:2013sra,Cristoforetti:2014gsa, 
  Alexandru:2015xva,Alexandru:2015sua,
  Fukuma:2017fjq,Fukuma:2019wbv} 
have been developed because of its mathematical rigor \cite{Witten:2010cx}. 
Along the line of such developments, 
the {\it tempered Lefschetz thimble method} (TLTM) was 
proposed \cite{Fukuma:2017fjq,Fukuma:2019wbv} 
as a versatile solution to the sign problem.%
\footnote{
 See \cite{Alexandru:2017oyw} for a similar idea. 
} 

As will be reviewed in the next section,  
the TLTM is a parallel-tempering algorithm, 
where the tempering parameter is set to be the flow time 
of the antiholomorphic gradient flow. 
This is to resolve the dilemma 
between the sign and ergodicity problems 
that becomes manifest when contributions from multiple thimbles are relevant.  
The validity of TLTM has been confirmed for various models, 
including the $(0+1)$-dimensional massive Thirring model \cite{Fukuma:2017fjq}, 
the Hubbard model away from half filling \cite{Fukuma:2019wbv} 
and a class of chiral matrix models (to be reported in another communication). 

As an algorithm to generate transitions on each flowed surface in TLTM, 
we have adopted the Metropolis algorithm in our previous study 
because of its simplicity. 
However, it is known that 
the Metropolis algorithm becomes less efficient 
than the Hybrid Monte Carlo (HMC) algorithm   
for systems including fermions with large degrees of freedom 
\cite{Duane:1987de,Creutz:1988wv}. 
Thus, the implementation of HMC on TLTM must give a useful framework 
for future computations of large-scale systems including fermions.

Our implementation of HMC on TLTM 
is based on the RATTLE algorithm \cite{Andersen:1983,Leimkuhler:1994} 
to realize molecular dynamics on flowed surfaces. 
RATTLE was first introduced to Lefschetz thimble methods 
in \cite{Fujii:2013sra} 
for molecular dynamics on a single Lefschetz thimble, 
and was generalized in \cite{Alexandru:2019}
for that on a flowed surface at finite flow time. 
Our achievement here is 
that HMC is implemented on TLTM 
so as to work within the parallel-tempering algorithm in TLTM, 
especially by developing an algorithm 
to handle zeros of fermion determinants 
in the course of the molecular-dynamics process.%
\footnote{
  Handling zeros of fermion determinants (in complex space $\bbC^N$) 
  is necessary 
  because they are reached with finite flow times 
  and configurations around them can be relevant to 
  observables under consideration. 
} 
To demonstrate that the implementation works correctly with high efficiency, 
we apply it to the Hubbard model away from half filling 
with small degrees of freedom 
($N=20$ for an $N_s=2\times 2$ spatial lattice 
with $N_\tau = 5$ imaginary time steps), 
for which the TLTM is known to work correctly 
with the Metropolis algorithm \cite{Fukuma:2019wbv}.%
\footnote{
 See 
 \cite{Mukherjee:2014hsa,Tanizaki:2015rda,Ulybyshev:2019a,Ulybyshev:2019b} 
 for related work on the application of Lefschetz thimble methods 
 to the Hubbard model.
} 
We show that our new algorithm gives results 
that agree nicely with exact values, 
and the computational cost to obtain an independent configuration 
is reduced to about 30\% of that of the Metropolis algorithm. 
We expect that greater efficiency will be gained for larger degrees of freedom. 

This paper is organized as follows. 
In section \ref{sec:preparations}, 
we first review the basics of TLTM. 
Then, after a short explanation on our convention, 
we give an outline for the implementation of HMC on TLTM. 
We in section \ref{sec:MD} explain a general theory 
for molecular dynamics on flowed surfaces, 
and in section \ref{sec:HMC_partial} 
give an explicit algorithm to implement HMC on TLTM. 
The algorithm is applied to the Hubbard model 
in section \ref{sec:results_analysis}, 
and is shown to work correctly 
with reduced computational costs compared with the Metropolis algorithm. 
Section \ref{sec:conclusion} is devoted to conclusion and outlook.

\section{Preparations}
\label{sec:preparations}

In this section, 
we first review the basics of TLTM. 
Then, after a short explanation on our convention, 
we give an outline for the implementation of HMC on TLTM. 

\subsection{Tempered Lefschetz thimble method (review)}
\label{sec:TLTM}

Let $\bbR^N=\{x\}$ be a configuration space 
of $N$-dimensional real variable $x=(x^i)$ $(i=1,\ldots,N)$, 
and $S(x)$ the action. 
Our main concern is to estimate the expectation value 
of an observable $\mathcal{O}(x)$, 
\begin{align}
 \langle \mathcal{O}(x) \rangle
 \equiv \frac{
 \int_{\bbR^N} dx\,e^{-S(x)}\,\mathcal{O}(x)
 }{
 \int_{\bbR^N} dx\,e^{-S(x)}
 }. 
\label{vev1}
\end{align}
In this paper, 
we always assume that both $e^{-S(z)}$ and $e^{-S(z)}\,\mathcal{O}(z)$ 
are entire functions over $\bbC^N$.%
\footnote{
  We also assume that 
  there is no multimodal problem on the original configuration space 
  $\bbR^N\,(\subset \bbC^N)$ 
  with respect to ${\rm Re}\,S(x)$. 
} 
Then, due to Cauchy's theorem in higher dimensions, 
the integrals in \eqref{vev1} do not change under continuous deformations 
of the integration region from $\bbR^N$ to $\Sigma$ 
with the boundary at $|x|\to\infty$ kept fixed:
\begin{align}
 \langle \mathcal{O}(x) \rangle 
 = \frac{\int_{\Sigma} d  z\, e^{-S(z)}\,\mathcal{O}(z)}
 {\int_{\Sigma} d z\, e^{-S(z)}}.
\label{LT1}
\end{align}
The sign problem will then get much reduced 
if $\textrm{Im}\, S(z)$ is almost constant on $\Sigma$. 

The Markov chain Monte Carlo (MCMC) calculation of 
\eqref{LT1} can be performed as follows. 
First, we decompose the complex measure $dz\equiv\prod_{i=1}^N dz^i$ 
to the modulus $|dz|$ and the phase $e^{i\varphi(z)}$, 
\begin{align}
 dz=|dz|\,e^{i\varphi(z)},
\end{align}
and rewrite $dz\,e^{-S(z)}$ as%
\footnote{
  Note that $|dz|$ agrees with the volume element 
  associated with the induced metric on $\Sigma$ in $\bbR^{2N}\,(=\bbC^N)$. 
} 
\begin{align}
 dz\,e^{-S(z)} = |dz|\,e^{-{\rm Re}\,S(z)}\,e^{i\theta(z)}
 \quad
 (e^{i \theta(z)} \equiv e^{i \varphi(z)} e^{-i\,\textrm{Im}\,S(z)} ),
\end{align}
from which \eqref{LT1} 
will be written as a ratio of reweighted integrals on $\Sigma$: 
\begin{align}
 \langle \mathcal{O}(x) \rangle
 = \frac{ \int_{\Sigma} dz \, e^{-S(z)}\,\mathcal{O}(z) }
 { \int_{\Sigma} dz \, e^{-S(z)} } 
 =\frac{
 \bigl\langle e^{i \theta(z)} 
 \mathcal{O}(z)
 \bigr\rangle_{\Sigma}}
 {\bigl\langle e^{i \theta(z)} 
 \bigr\rangle_{\Sigma}} 
\label{LT1a}
\end{align}
with
\begin{align}
 \langle f(z)\rangle_\Sigma 
 \equiv 
 \frac{1}{Z_\Sigma}
 \int_{\Sigma} |dz|\,e^{-\textrm{Re}\,S(z)} f(z)
 \quad \big(Z_\Sigma=\int_\Sigma |dz|\,e^{-{\rm Re}\,S(z)}\bigr). 
\label{measure}
\end{align}
We then generate a sample $\{z^{(k)}\}_{k=1,\ldots,N_{\rm conf}}$ 
from the distribution $e^{-{\rm Re}\,S(z)}/Z_\Sigma$,%
\footnote{
  When we consider probability densities $p({z})$ at ${z}\in\Sigma$, 
  they are always with respect to the measure $|d{z}|$. 
  The measure $|d{z}|$ 
  will also be written as $(d{z})_\parallel$ in later discussions.
  A transition from $p({z})$ to $\tilde{p}({z})$ 
  with transition probability $P({z}'|{z})$ is then expressed as 
  \begin{align}
   \tilde{p}({z}') = \int_\Sigma |d{z}|\,P({z}'|{z})\,p({z})
   \quad ({z}'\in\Sigma).
  \nonumber
  \end{align}
} 
and estimate $\langle f(z) \rangle_\Sigma$ as a sample average, 
\begin{align}
 \langle f(z)\rangle_\Sigma \approx 
 \frac{1}{N_{\rm conf}}\,\sum_{k=1}^{N_{\rm conf}} f(z^{(k)})
 \equiv \overline{f(z)}, 
\end{align}
from which $\langle \mathcal{O}(x) \rangle$ is estimated as%
\footnote{
  The statistical errors of  
  $\overline{f(z)}$ and the ratio $\bar{\mathcal{O}}$ 
  will be estimated using the binning-Jackknife method 
  (with autocorrelations taken into account). 
} 
\begin{align}
 \langle \mathcal{O}(x) \rangle 
 =\frac{
 \bigl\langle e^{i \theta(z)} 
 \mathcal{O}(z)
 \bigr\rangle_{\Sigma}}
 {\bigl\langle e^{i \theta(z)} 
 \bigr\rangle_{\Sigma}}
 \approx
 \frac{\overline{e^{i \theta(z)}\,\mathcal{O}(z)}}
 {\overline{e^{i\theta(z)}}}
 \equiv \bar{\mathcal{O}}.
\label{ratio}
\end{align}

In a class of Lefschetz thimble methods 
(see, e.g., \cite{Alexandru:2015xva,Alexandru:2015sua,
Fukuma:2017fjq,Fukuma:2019wbv}), 
continuous deformations of integration region are made 
according to the antiholomorphic flow equation: 
\begin{align}
 \dot{z}_t^i &= [\partial_i S(z_t)]^\ast,
 \quad
 z^i_{t=0} = x^i, 
\label{flow_z}
\end{align}
which defines a map from $x\in\bbR^N$ to $z=z_t(x)\in\bbC^N$. 
Since $(d/dt)\,S(z_t)=|\partial_i S(z_t)|^2\geq 0$, 
the real part ${\rm Re}\,S(z_t)$ always increase along the flow 
except at critical points $z_\sigma$ 
(where $\partial_i S(z_\sigma)=0$), 
while the imaginary part ${\rm Im}\,S(z_t)$ is kept constant. 
In the limit $t\to \infty$, 
$\Sigma_t\equiv z_t(\bbR^N)$ will approach a union of Lefschetz thimbles, 
on each of which ${\rm Im}\,S(z)$ is constant, 
and thus the sign problem is expected to disappear there 
(except for a possible residual and/or global sign problem).%
\footnote{
 In the case when the action diverges at some points in $\bbC^N$ 
 (such as zeros of the fermion determinant),
 $\Sigma_t$ should be understood to represent $z_t(\bbR^N)$ 
 with these points removed.
\label{fn:removal}
} 
However, for large $t$ there arises a new problem, 
multimodal (ergodicity) problem, 
because the potential barriers between different thimbles 
become infinitely high 
as $t$ increases.

In the tempered Lefschetz thimble method (TLTM) 
\cite{Fukuma:2017fjq,Fukuma:2019wbv}, 
we resolve the dilemma between the sign problem (severe at small flow times) 
and the ergodicity problem (severe at large flow times) 
by tempering the system with the flow time.%
\footnote{
 As a tempering algorithm, 
 we adopt the parallel tempering (also called the replica exchange MCMC method) 
 \cite{Swendsen1986,Geyer1991,Hukushima1996} 
 because then we need not specify the probability weight factors 
 at various flow times 
 and because most of relevant steps can be done in parallel processes. 
} 
The algorithm consists of three steps. 
(1) First, we introduce a set of configuration spaces, 
$\{\Sigma_{t_a}\}$ $(a=0,1,\ldots,A)$,
with $t_0=0 < t_1 < \cdots < t_A=T$.
We often call $\Sigma_{t_a}$ the $a$-th replica. 
Here, a possible criterion for choosing the maximum flow time $T$ is 
that the sign average $|\langle e^{i \theta(z)}\rangle_{\Sigma_T}|$
is $O(1)$ without tempering. 
(2) We then construct a Markov chain 
that drives the enlarged system 
$\Sigma_{\rm tot}\equiv
\Sigma_{t_0}\times \Sigma_{t_1}\times\cdots\times\Sigma_{t_A}
=\{\vec z=(z_a)\}$
to global equilibrium 
with the distribution 
$p_{\rm eq}(\vec z)\propto\prod_a \exp[-{\rm Re}\,S(z_a)]$. 
(3) After the system is well relaxed to global equilibrium, 
we estimate the expectation value on $\Sigma_{t_a}$ 
[see \eqref{ratio}]
by using the subsample at replica $a$, 
$\{z_a^{(k)}\}_{k=1,2,\ldots,N_{\rm conf}}$, 
that is retrieved from the total sample 
$\{\vec z^{(k)}
=(z_0^{(k)},z_1^{(k)},\ldots,z_A^{(k)})\}_{k=1,2,\ldots,N_{\rm conf}}$: 
\begin{align}
 \frac{
 \bigl\langle e^{i \theta(z_a)} 
 \mathcal{O}(z_a)
  \bigr\rangle_{\Sigma_{t_a}}}
 {\bigl\langle e^{i \theta(z_a)} 
 \bigr\rangle_{\Sigma_{t_a}}}
 \thickapprox
 \frac{\overline{e^{i \theta(z_a)}\,\mathcal{O}(z_a)}}
 {\overline{e^{i\theta(z_a)}}}
 \equiv \bar{\mathcal{O}}_a.
\label{estimate}
\end{align}
Since the left-hand side of \eqref{estimate} 
is independent of $a$ due to Cauchy's theorem, 
the ratio $\bar{\mathcal{O}}_a$ at large $a$'s 
(where the sign problem is relaxed)
should yield the same value 
within the statistical error margin 
if the system is well in global equilibrium. 
Conversely, the requirement of $a$-independence 
ensures the sample to be in global equilibrium 
with a sufficient sample size 
(together with the correctness of the employed numerical method), 
and is the basis of the following algorithm for precise estimation 
\cite{Fukuma:2019wbv}. 
First, we continue the sampling 
until we find some range of $a$, 
in which 
(i) $\bigl|\overline{e^{i\theta(z_a)}}\bigr|$ 
are well above $1/\sqrt{2N_{\rm conf}}$ 
(the values for the uniform distribution of phases)
and (ii) $\bar{\mathcal{O}}_a$ take the same value 
within the statistical error margin.  
Then, we estimate $\langle \mathcal{O} \rangle$ 
by using the $\chi^2$ fit (using covariance) of
$\{\bar{\mathcal{O}}_a\}$ in this region
with a constant function of $a$. 
Global equilibrium and the sufficiency of the sample size 
are checked by looking at the optimized value of 
$\chi^2/{\rm DOF}$. 

\subsection{Real representation for complex variables}
\label{sec:real_representation}

In the following sections, 
we mainly use the real representation for complex variables, 
where a point $z=(z^i)=x + i y \in \bbC^N$ $(i=1,\ldots,N)$
is expressed by ${z}=({z}^I)\equiv(x,y)^T \in \bbR^{2N}$ $(I=1,\ldots,2N)$.%
\footnote{
  We use the same symbol for both representations 
  to avoid introducing many redundant symbols. 
  Which representation is used should be obvious from the context, 
  and otherwise it will be clearly stated. 
} 
Accordingly, a complex column vector $v=(v^i)=v_R + i v_I\in \bbC^N$ 
will be written as a real vector  
$v=(v^I)\equiv (v_R,v_I)^T \in\bbR^{2N}$.
Under this identification, 
an $N\times N$ complex matrix $A=(A_{ij})=A_R + i A_I$ 
will be given as a $2N\times 2N$ matrix,
\begin{align}
 A = (A_{IJ}) \equiv \left(
 \begin{array}{cc}
  A_R & - A_I \\
  A_I & A_R \\
 \end {array}
 \right). 
\label{A_real}
\end{align}
We write the multiplication of $i$ (imaginary unit) 
and the complex conjugation, respectively, as
\begin{align}
 {\hat{i}} \equiv \left(
 \begin{array}{cc}
  0 & -1_N \\
  1_N & 0 \\
 \end {array}
 \right),
 \quad
 {\hat{C}} \equiv \left(
 \begin{array}{cc}
  1_N & 0 \\
  0 & -1_N \\
 \end {array}
 \right),
\end{align}
and introduce the projectors to the real and imaginary parts, respectively,
as
\begin{align}
 \widehat{\rm Re} \equiv \frac{1}{2}\,\bigl(1+\hat{C}\bigr)
 = \left(
 \begin{array}{cc}
  1_N & 0 \\
  0 & 0 \\
 \end {array}
 \right),
 \quad
 \widehat{\rm Im} \equiv \frac{1}{2}\,\bigl(1-\hat{C}\bigr)
 = \left(
 \begin{array}{cc}
  0 & 0 \\
  0 & 1_N \\
 \end {array}
 \right).
\end{align}

\subsection{Outline for the implementation of HMC on TLTM}
\label{sec:outline}

We will often abbreviate flowed surfaces $\Sigma_{t_a}=\{{z}_a\}$ 
as $\Sigma_a$ ($a=0,1,\ldots,A$) to simplify expressions. 
In the parallel tempering, 
the total configuration space is given by 
\begin{align}
 \Sigma_{\rm tot}\equiv \Sigma_0\times\cdots\times\Sigma_A
= \{ \vec{z}=({z}_a)\},
\end{align}
which we regard as a complex of playgrounds with $A+1$ zones 
for the same number of molecules, 
where each molecule moves around from a zone to another zone
under the condition that 
any two molecules cannot be in the same zone.%
\footnote{ 
 Note that the index $a$ labels the zones, not the molecules.
} 
In order to implement an HMC algorithm on the tempered system, 
we introduce to each replica $\Sigma_a$  
the phase space $T^\ast\Sigma_a=\{\zeta_a=({z}_a,\pi_a)\}$ 
and the Hamiltonian 
\begin{align}
 H_a(\zeta_a) \equiv \frac{\pi_a^2}{2 M_a} + V({z}_a),
\end{align}
where $\pi_a^2/2 M_a \equiv (1/2)\,(M_a^{-1})^{IJ}\pi_{a,I}\,\pi_{a,J}$ 
and $V({z}_a)\equiv{\rm Re}\,S({z}_a)$. 
$M_a$ is constant 
and will be set to be $(M_a)_{IJ}=\sigma_a^2\,\delta_{IJ}$. 

We construct a molecular dynamics 
on each phase space $T^\ast\Sigma_a$ 
(to be explained in detail in the next section), 
that defines a one-body motion from $\zeta_a\in T^\ast\Sigma_a$ 
to $\Phi_a(\zeta_a)\in T^\ast\Sigma_a$. 
$\Phi_a$ will be designed such that 
it is volume-preserving and reversible, 
and thus the transition probability%
\footnote{
  The diagonal elements are determined automatically 
  by the probability conservation 
  $\int d\zeta'_a\,\mathbb{P}_a^{(1)}(\zeta'_a|\zeta_a)=1$. 
  This comment will be applied to similar expressions in what follows.
} 
\begin{align}
 \mathbb{P}^{(1)}_a(\zeta'_a|\zeta_a)
 \equiv 
 \min\bigl(1,e^{-H_a(\zeta'_a)+H_a(\zeta_a)}\bigr)\,
 \delta(\zeta'_a-\Phi_a(\zeta_a))
 \quad (\zeta'_a\neq\zeta_a)
\label{W(1)}
\end{align}
satisfies the relation 
\begin{align}
 \mathbb{P}^{(1)}_a(\zeta'_a|\zeta_a)\,e^{-H_a(\zeta_a)}
 = \mathbb{P}^{(1)}_a(\zeta_a^T|\zeta^{\prime\,T}_a)\,e^{-H_a(\zeta'_a)}
 \quad (\zeta'_a\neq\zeta_a),
\label{W(1)_det}
\end{align}
where $\zeta^T\equiv ({z},-\pi)$ for $\zeta=({z},\pi)$ 
and we have used the fact $H_a(\zeta^T_a)=H_a(\zeta_a)$.

We then define the transition probability on each replica:%
\footnote{
 $d\pi_a$ is the volume element of $T^\ast_{z_a}\Sigma_a$ 
 and will be denoted by $(d\pi_a)_\parallel$ 
 when $\pi_a$ is an element in $T^\ast_{z_a}\bbR^{2N}$.
} 
\begin{align}
 P^{(1)}_a({z}'_a|{z}_a)
 \equiv 
 c_a\,\int d\pi'_a d\pi_a\,
 \mathbb{P}^{(1)}_a({z}'_a,\pi'_a| {z}_a,\pi_a)\,
 e^{-\pi_a^2/2M_a}
\end{align}
with $c_a\equiv \bigl[\int d\pi_a\,e^{-\pi_a^2/2M_a}\bigr]^{-1}$. 
$P^{(1)}_a$ satisfies the following detailed balance condition:
\begin{align}
 P^{(1)}_a({z}'_a|{z}_a)\,e^{-V({z}_a)}
 = P^{(1)}_a({z}_a|{z}'_a)\,e^{-V({z}'_a)}.
\end{align}
We also introduce a two-body evolution 
that maps $({z}_a,{z}_b)\in\Sigma_a\times\Sigma_b$
to $({z}'_a,{z}'_b)\in\Sigma_a\times\Sigma_b$
with the probability 
$P^{(2)}_{ab}({z}'_a,{z}'_b|{z}_a,{z}_b)$ 
satisfying the relation
\begin{align}
 P^{(2)}_{ab}({z}'_a,{z}'_b|{z}_a,{z}_b)\,
 e^{-V({z}_a)-V({z}_b)}
 = P^{(2)}_{ab}({z}_a,{z}_b|{z}'_a,{z}'_b)\,
 e^{-V({z}'_a)-V({z}'_b)}.
\label{two-body}
\end{align}
By combining $P^{(1)}_a$ and $P^{(2)}_{ab}$, 
one can construct a Markov chain 
such that its transition probability 
$P_{\rm tot}(\vec{z}'|\vec{z})$ 
gives the desired equilibrium distribution 
$p_{\rm eq}(\vec{z})\propto \prod_a e^{-V({z}_a)}$.

In the following sections, 
we define a molecular dynamics on each replica 
(section \ref{sec:MD}) 
and then give an explicit algorithm 
for HMC on TLTM (section \ref{sec:HMC_partial}). 

\section{Molecular dynamics on flowed surfaces}
\label{sec:MD}

In this section, we first give a brief review 
of molecular dynamics on a general constrained surface $\Sigma$, 
and then discuss molecular dynamics on a flowed surface $\Sigma=\Sigma_t$ 
that is obtained as a time slice 
from the antiholomorphic gradient flow with flow time $t$.

\subsection{Molecular dynamics on a general constrained surface}
\label{sec:MD_general}

Let $\Sigma$ be an $m$-dimensional surface in $\bbR^{2N} \,(=\bbC^N)$, 
which we assume is given by a set of constraint equations%
\footnote{
 Later we will set $m=N$.
} 
\begin{align}
 \phi^r({z}) = 0 \quad (r=1,\ldots,2N-m).
\end{align}
At point ${z}\in\Sigma$, 
we choose a basis of the tangent space $T_{{z}}\Sigma$ 
and denote it by $E_{\alpha}=(E_{\alpha}^I)$ $({\alpha}=1,\ldots,m)$, 
from which we define the metric 
$g_{{\alpha}{\beta}}\equiv \delta_{IJ}\,E^I_{\alpha} E^J_{\beta}$. 
We also introduce a basis $F_r=(F_r^I)$ 
of the normal space $N_{{z}}\Sigma$.  

\begin{figure}[ht]
\centering
\includegraphics[width=70mm]{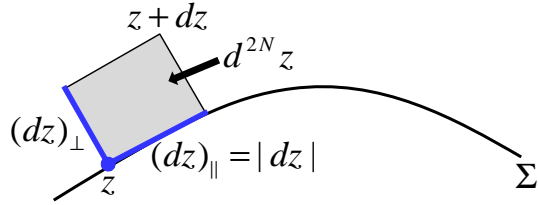} 
\caption{
\label{fig:volume}
 Orthogonal decomposition of $d^{2N}z$.
}
\end{figure}

Let $T^\ast \bbR^{2N} =\{\zeta=({z},\pi)\}$ 
be the phase space on $\bbR^{2N}$ 
with a separable Hamiltonian of the form 
\begin{align}
 H(\zeta) = H({z},\pi) = \frac{1}{2}\,(M^{-1})^{IJ} \pi_I\pi_J + V({z}),
\end{align}
where the positive symmetric mass matrix $M=(M_{IJ})$ 
is assumed to be constant.%
\footnote{
  We assume that $M$ satisfies the condition 
  $M_{IJ} E^I_{\alpha}\, F^J_r =0$,
  which ensures that $(M^{-1})^{IJ} \pi_J$ is on the tangent space $T_{z}\Sigma$ 
  for $\pi=(\pi_I)\in T^\ast_{z}\Sigma$.
} 
A motion on $\Sigma$ defines a motion in the reduced phase space 
\begin{align}
 T^\ast \Sigma \equiv \{\zeta=({z},\pi)\in T^\ast \bbR^{2N} \,|\, 
 \phi^r({z}) =0,~ 
 (M^{-1} \pi)^I\, \partial_{{z}^I}\phi^r({z})=0\}.
\end{align}
The symplectic structure of $T^\ast\Sigma$ is defined 
by the induced symplectic form
\begin{align}
 \omega \equiv d\pi_I\wedge d{z}^I|_{T^\ast\Sigma},
\label{omega}
\end{align}
from which we define the volume element $dV$ on $T^\ast \Sigma$ by 
\begin{align}
 dV \equiv \frac{\omega^m}{m!}. 
\label{dV1}
\end{align}
When $d^{2N}\!{z}\equiv \prod_I d{z}^I$
and $d^{2N}\!\pi\equiv \prod_I d\pi_I$ 
are orthogonally decomposed as (see Fig.~\ref{fig:volume})
\begin{align}
 d^{2N}\!{z} = (d{z})_\parallel\,(d{z})_\perp,
 \quad
 d^{2N}\!\pi = (d\pi)_\parallel\,(d\pi)_\perp,
\end{align}
one can easily show that 
\begin{align}
 dV = (d{z})_\parallel\,(d\pi)_\parallel.
\label{dV2}
\end{align}
\noindent
Note that $(dz)_\parallel = |dz|$ 
and $(d\pi)_\parallel$ corresponds to $d\pi$ in subsection \ref{sec:outline}.
If we introduce local coordinates ${\xi}=({\xi}^{\alpha})$ on $\Sigma$, 
then we can choose the basis of $T_{z}\Sigma$ to be 
$E^I_{\alpha}=\partial {z}^I/\partial {\xi}^{\alpha}$. 
It is convenient to define the projected components $\eta_{\alpha}$
for arbitrary momentum $\tilde\pi=(\tilde\pi_I)\in T^\ast_{z}\bbR^{2N}$: 
\begin{align}
 \eta_{\alpha} \equiv \tilde\pi_I\,E^I_{\alpha}. 
\end{align}
One can easily show that 
\begin{align}
 \omega = d\eta_{\alpha}\wedge d{\xi}^{\alpha},
\end{align}
and thus the volume element can be expressed as%
\footnote{
  With the local coordinates, 
  each factor in \eqref{dV2} is written as 
  $(d{z})_\parallel=|dz|=\sqrt{g}\,\prod_{\alpha} d\xi^{\alpha}$ 
  and $(d\pi)_\parallel=(1/\sqrt{g})\,\prod_{\alpha} d\eta_{\alpha}$
  with $g=\det(g_{{\alpha}{\beta}}(\xi))$. 
} 
\begin{align}
 dV = \prod_{\alpha} d{\xi}^{\alpha} d\eta_{\alpha}.
\label{dV3}
\end{align}
One can also show that the projection of $\tilde\pi$ 
to $\pi=\tilde\pi_\parallel$ 
is given by
\begin{align}
 \pi_{I} = \eta_{\alpha}\,E^{\alpha}_I
 = \tilde\pi_J\,\mathcal{P}^J_{~I},
\end{align}
where $E^{\alpha}_I \equiv g^{{\alpha}{\beta}}\,\delta_{IJ}\,E^J_{\beta}$ 
[$(g^{{\alpha}{\beta}})\equiv (g_{{\alpha}{\beta}})^{-1}$]
and $\mathcal{P}^J_{~I} \equiv E^J_{\alpha} E^{\alpha}_I$.

In the continuous language, 
a motion $\zeta(s)=({z}(s),\pi(s))$ in $T^\ast\Sigma$ 
is described by the following equations 
with Lagrange multipliers $\lambda_r$: 
\begin{align}
 \partial_s{z}^I &= \partial_{\pi_I} H = (M^{-1}\pi)^I,
\label{zetadot}
\\
 \partial_s\pi_I &=
 {} -\partial_{{z}^I} H - \lambda_r\,\partial_{{z}^I} \phi^r({z})
 ={} -\partial_{{z}^I} V({z}) - \lambda_r \,\partial_{{z}^I} \phi^r({z}),
\label{pidot}
\\
 0 &= \phi^r({z}),
\label{phi}
\\
 0 &= (M^{-1}\pi)^I\, \partial_{{z}^I}\phi^r({z}). 
\label{tangent}
\end{align}
Note that \eqref{tangent} 
(obtained by taking the derivative of \eqref{phi} with respect to $s$) 
means that the velocity $\partial_s{z}=M^{-1}\pi$ is tangent to $\Sigma$. 
Equations \eqref{zetadot}--\eqref{tangent} have the following properties: 
(1) \underline{symplecticity}:  
The induced symplectic form \eqref{omega} 
does not change under the motion, $\partial_s{\omega}=0$, 
and thus the volume element $dV$ is preserved. 
(2) \underline{reversibility}: 
For any motion $({z},\pi) \stackrel{s}{\to} ({z}',\pi')$ 
obtained by integrating \eqref{zetadot}--\eqref{tangent}, 
its time-reversed motion $({z}',-\pi') \stackrel{s}{\to} ({z},-\pi)$ 
is also a solution. 
(3) \underline{energy conservation}: 
$H({z}',\pi')=H({z},\pi)$. 

RATTLE \cite{Andersen:1983,Leimkuhler:1994} 
is a discrete version of the above molecular dynamics, 
and generates a one-step motion from $({z},\pi)\in T^\ast \Sigma$ 
to $({z}',\pi')\equiv\Phi_{\Delta s}({z},\pi)\in T^\ast \Sigma$ 
as follows: 
\begin{align}
 \pi_{1/2} &= \pi -\frac{\Delta s}{2}\,\partial V({z})
 -\frac{\Delta s}{2}\,\lambda^{(1)}_r\,\partial\phi^r({z}),
\label{RATTLE1}
\\
 {z}' &= {z} + \Delta s\,(M^{-1}\,\pi_{1/2}),
\label{RATTLE2}
\\
 0 &= \phi^r({z}'),
\label{RATTLE3}
\\
 \pi' &= \pi_{1/2} - \frac{\Delta s}{2}\,\partial V({z}')
 - \frac{\Delta s}{2}\,\lambda^{(2)}_r\,\partial\phi^r({z}'),
\label{RATTLE4}
\\
 0 &= (M^{-1}\pi')\cdot\partial\phi^r({z}'),
\label{RATTLE5}
\end{align}
where $\Delta s$ is the step size 
and $\partial V({z}) \equiv (\partial_{{z}^I} V({z}))$. 
Note that there appear two Lagrange multipliers. 
$\lambda^{(1)}_r$ is determined 
so that the new configuration ${z}'$ is on $\Sigma$
[${z}'\in\Sigma$, eq.~\eqref{RATTLE3}], 
while $\lambda^{(2)}_r$ is determined 
so that $\pi'$ is in the tangential direction  
[$\pi'\in T^\ast_{{z}'} \Sigma$, eq.~\eqref{RATTLE5}]. 
One can easily check that 
$\Phi_{\Delta s}$ is symplectic 
($d\pi'\wedge d{z}'|_{T^\ast\Sigma}=d\pi\wedge d{z}|_{T^\ast\Sigma}$)  
and reversible 
[if $({z}',\pi') = \Phi_{\Delta s}({z},\pi)$ 
then $({z},-\pi) = \Phi_{\Delta s}({z}',-\pi')$ 
with the interchange of $\lambda^{(1)}$ and $\lambda^{(2)}$], 
and preserves the energy to second order:
\begin{align}
 H({z}',\pi') = H({z},\pi) + O(\Delta s^3). 
\end{align}

\begin{figure}[ht]
\centering
\includegraphics[width=110mm]{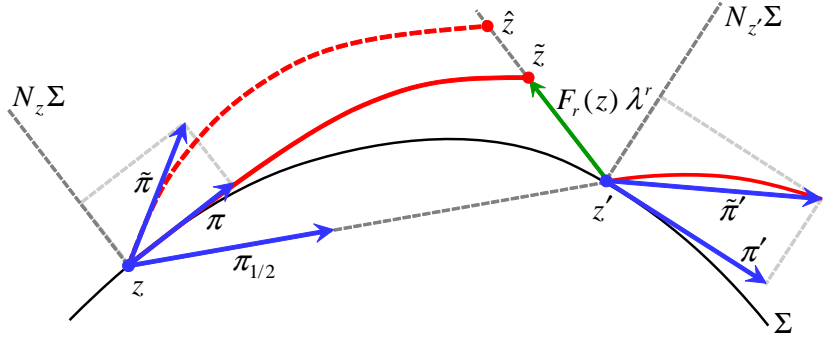} 
\caption{
\label{fig:RATTLE}
 RATTLE process from $\zeta=({z},\pi)\in T^\ast_z\Sigma$ 
 to $\zeta'=({z}',\pi')\in T^\ast_{z'}\Sigma$.
}
\end{figure}

We need to rewrite the above RATTLE process 
when, as in Lefschetz thimble methods \cite{Fujii:2013sra}, 
we do not know explicit functional forms 
of the constraint functions $\phi^r({z})$ 
except for the bases of the tangent and the normal spaces at ${z}$ 
[$E_{\alpha}=(E_{\alpha}^I)$ and $F_r=(F_r^I)$, respectively]. 
This rewriting can be done in the following way 
by using the orthogonal projector 
$\mathcal{P}({z}) = (\mathcal{P}^I_{~J}({z}))$ 
from $T_{z}^\ast\bbR^{2N}$ to $T_{z}^\ast\Sigma$ 
(see Fig.~\ref{fig:RATTLE}):%
\footnote{
  We owe much of the discussions for RATTLE 
  on a flowed surface to \cite{Alexandru:2019}, 
  where a generalization is made from the RATTLE 
  on a single Lefschetz thimble \cite{Fujii:2013sra}. 
} 
\noindent
\begin{enumerate}
\item
For a given $\zeta=({z},\pi)\in T^\ast \Sigma$, 
we set 
\begin{align}
 \tilde{{z}} \equiv {z} + \Delta s\, (M^{-1}\pi) 
 - \frac{\Delta s^2}{2}\,M^{-1}\,\partial V({z}).
\label{newRATTLE1}
\end{align}

\item
We find $\lambda^r$ such that 
\begin{align}
 {z}' \equiv \tilde{{z}} - F_r({z})\,\lambda^r \in \Sigma. 
\label{newRATTLE2}
\end{align}

\item
We define
\begin{align}
 \pi_{1/2} \equiv \frac{1}{\Delta s}\,M ({z}'-{z}),
\label{newRATTLE3}
\end{align}
and set
\begin{align}
 \tilde{\pi}' \equiv \pi_{1/2} - \frac{\Delta s}{2}\, \partial V({z}'). 
\label{newRATTLE4}
\end{align}

\item
We project $\tilde\pi'$ on $T^\ast_{{z}'}\Sigma$: 
\begin{align}
 \pi' \equiv \tilde{\pi}'\,\mathcal{P}({z}').
\label{newRATTLE5}
\end{align}
\end{enumerate}
Note that $\lambda^r$ must be of $O(\Delta s^2)$.%
\footnote{
 Note that 
 the replacement of $\pi$ in \eqref{newRATTLE1} by $\tilde\pi$ 
 in Fig.~\ref{fig:RATTLE} (and thus $\tilde{z}$ by $\hat{z}$)
 can be totally absorbed by a shift of the Lagrange multiplier $\lambda^r$, 
 without changing the location of ${z}'$. 
 With this replacement, 
 we can rewrite the above steps 
 as a procedure to obtain a new pair $({z}',\tilde\pi')$ from $({z},\tilde\pi)$ 
 as in Fig.~\ref{fig:RATTLE}. 
 The projection (Step 4) is then required 
 only at the final step of molecular-dynamics evolution. 
 Note that the Lagrange multiplier will then become of $O(\Delta s)$. 
} 

\subsection{Molecular dynamics on a flowed surface $\Sigma_t$}
\label{sec:MD_Sigma_t}

We now apply the formalism developed in the previous subsection 
to a flowed surface $\Sigma=\Sigma_t$ in the TLTM, 
where $m\,(={\rm dim}\,\Sigma_t)$ is $N$.%
\footnote{
  We will label the constraints also with ${\alpha}\,(=1,\ldots,N)$ 
  instead of $r$. 
} 
The potential is given by $V({z}) = \textrm{Re}\,S({z})$. 
A crucial point here is that, 
although we do not know explicit functional forms of $\phi^{\alpha}({z})$, 
there is a one-to-one correspondence 
between the points ${z}=({z}^I)\in\Sigma_t$ 
and those $x=(x^{\alpha})\in\bbR^N$ 
with the relation $z=z_t(x)$. 
Furthermore, the bases of the tangent and normal spaces 
at ${z}={z}_t(x)$ can be given explicitly as 
\begin{align}
 E_{\alpha}^I(x) \equiv \partial{z}_t^I(x)/\partial x^{\alpha},
 \quad
 F_{\alpha}^I(x) \equiv (\,{\hat{i}}\,E_{\alpha}(x))^I \quad (I=1,\ldots,2N),
\end{align}
whose complex representations are  
$E^i_{\alpha} = \partial z_t^i(x)/\partial x^{\alpha}$ and 
$F^i_{\alpha} = i\,E^i_{\alpha}$ 
$(i=1,\ldots,N)$ \cite{Fujii:2013sra}. 
Note that $(E^i_{\alpha}(x))$ is nothing but 
the (complex-valued) Jacobian matrix 
$J(x)= J_t(x)= (\partial z_t^i(x)/\partial x^{\alpha})$, 
that obeys the following differential equation 
in the complex representation 
\cite{Alexandru:2015xva} 
(see also footnote 2 of \cite{Fukuma:2017fjq}): 
\begin{align}
 \dot{J}_t &= [ H(z_t)\cdot J_t]^\ast,
 \quad
 J_{t=0} = 1_N
\label{flow_J}
\end{align}
with $H(z)\equiv (\partial_i \partial_j S(z))$.
The real representation of $J(x)=J_{R}(x)+i J_{I}(x)$ 
is in turn given by%
\footnote{
  With \eqref{flow_J} and \eqref{J_real}
  the orthogonality between $E_\alpha=(E_\alpha^I)$ 
  and $F_\alpha=(F_\alpha^I)$ can be shown as follows 
  \cite{Fujii:2013sra}.
  We first note that their inner products
  $E_\alpha\cdot F_\beta\equiv\delta_{IJ}\,E^I_\alpha F^J_\beta$
  can be written as $-{\rm Im}\,(J_t^\dag J_t)_{\alpha\beta}$, 
  and that they do not depend on $t$ 
  due to the first equation of \eqref{flow_J}. 
  We then conclude that $E_\alpha\cdot F_\beta$ must vanish 
  due to the initial condition $J_{t=0}=1_N$, 
  for which $-{\rm Im}\,(J_t^\dag J_t)|_{t=0}=0$. 
} 
\begin{align}
 J(x) = (J^I_{\,A}(x)) \equiv \left(
 \begin{array}{cc}
  (J_R(x))^i_{~{\alpha}} & - (J_I(x))^i_{~{\alpha}} \\
  (J_I(x))^i_{~{\alpha}} & (J_R(x))^i_{~{\alpha}} \\
 \end {array}
 \right)
 = (E_{\alpha}(x),\, F_{\alpha}(x))
 \quad (A=1,\ldots,2N). 
\label{J_real}
\end{align}

The condition that ${z}' \in \Sigma_t\subset \bbR^{2N}$ 
for a given ${z}={z}_t(x)$ 
[eq.~\eqref{newRATTLE2}]
is equivalent to  
that there be a vector $u=(u^{\alpha}) \in \bbR^N$ 
such that ${z}'$ can be written as ${z}'={z}_t(x+u)$. 
Together with the need to find $\lambda=(\lambda^{\alpha})\in\bbR^N$, 
our requirement can be expressed as the following $2N$ equations 
for $2N$ unknown variables $u^{\alpha},\lambda^{\alpha}$ $({\alpha}=1,\ldots,N)$: 
\begin{align}
 0 &= {z}_t^I (x+u) -\tilde{{z}}^I + F^I_{\alpha}({z})\,\lambda^{\alpha}
\nonumber
\\
 &= {z}_t^I (x+u) - {z}_t^I(x) - \Delta s\, (M^{-1}\pi)^I 
 + \frac{\Delta s^2}{2}\,(M^{-1}\,\partial V({z}))^I 
 + F^I_{\alpha}({z})\,\lambda^{\alpha}
\nonumber
\\
 &\equiv f^I(u,\lambda;\,x). 
\label{ul_eq}
\end{align}
This equation can be solved iteratively for 
\begin{align}
 w=(w^A) = \left(
 \begin{array}{c}
  u^{\alpha} \\
  \lambda^{\alpha}
 \end{array}\right)
\end{align}
with Newton's method. 
Namely, starting from an initial guess $w_0=(w^A_0)$, 
we obtain a sequence $w_k \to w_{k+1}=w_k + \Delta w$ 
by solving the linear equation
\begin{align}
 \frac{\partial f^I}{\partial w^A}\Bigr|_{w_k}\,\Delta w^A ={} -f^I(w_k). 
\label{Newton}
\end{align}
Here, from the explicit form of $f^I$, we find that 
\begin{align}
 \frac{\partial f^I}{\partial u^{\alpha}} 
 &= \frac{\partial z^I(x+u)}{\partial u^{\alpha}}
 =\frac{\partial z^I(x)}{\partial x^{\alpha}}\Bigr|_{x+u}
 =E_{\alpha}^I(x+u),
\\
 \frac{\partial f^I}{\partial \lambda^{\alpha}} &= F_{\alpha}^I(x),
\end{align}
and thus the recursive equation can be written as 
\begin{align}
 \bigl(E_{\alpha}(x+u_k),\, F_{\alpha}(x) \bigr) 
 \left(
 \begin{array}{c}
   \Delta u^{\alpha} \\
   \Delta \lambda^{\alpha} \\
 \end{array}
 \right)
 ={} - f(w_k), 
\label{recursion1}
\end{align}
or equivalently,
\begin{align}
  \left(
  \begin{array}{cc}
    J_R(x+u_k) & -J_I(x) \\
    J_I(x+u_k) & J_R(x) \\
  \end{array}
  \right)
  \left(
  \begin{array}{c}
    \Delta u \\
    \Delta \lambda \\
  \end{array}
  \right)
  ={} - f(w_k). 
\label{recursion2}
\end{align}

The linear equation \eqref{recursion2} can be solved in two ways. 
One is to directly obtain all the matrix elements of 
the Jacobian matrices $J(x)=J_R(x) + i\,J_I(x)$ 
and $J(x+u_k)=J_R(x+u_k) + i\,J_I(x+u_k)$ 
by numerically integrating \eqref{flow_z} and \eqref{flow_J} 
and then to obtain the solution $\Delta w = (\Delta u, \Delta \lambda)^T$  
with a direct method such as the LU decomposition. 
The other method is to use an iterative method 
such as GMRES \cite{Saad:1983} or BiCGStab \cite{Vorst:1990}
without calculating the matrix elements explicitly 
(as in \cite{Alexandru:2017lqr}). 
The reason why such a method is possible here
is that the left-hand side of \eqref{recursion2} can be rewritten 
as 
\begin{align}
  \left(
  \begin{array}{cc}
    J_R(x+u_k) & -J_I(x+u_k) \\
    J_I(x+u_k) & J_R(x+u_k) \\
  \end{array}
  \right)
  \left(
  \begin{array}{c}
    \Delta u \\
    0 \\
  \end{array}
  \right)
 +
  \hat{i}\,
  \left(
  \begin{array}{cc}
    J_R(x) & -J_I(x) \\
    J_I(x) & J_R(x) \\
  \end{array}
  \right)
  \left(
  \begin{array}{c}
    \Delta \lambda \\
    0 \\
  \end{array}
  \right), 
\end{align}
and each term can be evaluated 
by numerically integrating the following differential equations 
for $z_t=(z_t^I)$ and a vector $v_t=(v_t^I)$ (not for a matrix):
\begin{align}
 \dot{{z}}_t &= \hat{C}\, \partial\, {\rm Re}S({z}_t),
\label{flow_z2}
\\
 \dot{v}_t &= \hat{C}\, H(z_t)\,v_t
 = 
 \left(
 \begin{array}{cc}
  H_R({z}_t) & -H_I({z}_t) \\
  -H_I({z}_t) & -H_R({z}_t) \\
 \end{array}
 \right) v_t
\label{flow_v2}
\end{align}
with the initial conditions 
$z_0=x+u_k$ and $v_0=(\Delta u,0)^T$ 
or $z_0=x$ and $v_0 = (\Delta \lambda,0)^T$. 
Note that the complex representations 
of \eqref{flow_z2} and \eqref{flow_v2}
for $z_t=(z_t^i),\,v_t=(v_t^i) \in\bbC^N$ 
are given, respectively, by
\begin{align}
 \dot{z}_t^i = [\partial_i S(z_t)]^\ast,
 \quad
 \dot{v}_t^i = [H_{ij}(z_t)\,v_t^j]^\ast.
\label{flow_zv_complex}
\end{align}
The initial conditions are then given by 
$z_0^i=x^i+u_k^i$ and $v_0^i=\Delta u^i$
or $z_0^i=x^i$ and $v_0^i=\Delta \lambda^i$ 
for $\Delta u,\,\Delta\lambda\in\bbR^N$.

In the procedure given in subsection \ref{sec:MD_general}, 
one needs to use the projector 
$\mathcal{P}({z}) = (\mathcal{P}^I_{~J}({z}))$ 
that projects $\tilde\pi\in T^\ast_{z}\bbR^{2N}$ 
to $\pi=\tilde\pi\,\mathcal{P}({z})\in T^\ast_{z}\Sigma_t$ 
at $z=z_t(x)$, 
or equivalently, 
that projects $\tilde v\equiv M^{-1}\tilde\pi\in T_{z}\bbR^{2N}$ 
to $v\equiv M^{-1}\pi=\mathcal{P}(z)\,\tilde v\in T_{z}\Sigma_t$
at $z=z_t(x)$. 
The projection 
can also be given in two ways. 
When the matrix elements of $J(x)$ are known explicitly 
as in the direct method given in the previous paragraph, 
the matrix $\mathcal{P} = (\mathcal{P}^I_{~J}=E^I_{\alpha}\,E^{\alpha}_J)$ 
can also be calculated explicitly as 
\begin{align}
 \mathcal{P} = (E^I_{\alpha})\,(E^{\alpha}_J)
 =  
 (E^I_{\alpha},\,F^I_{\alpha}) 
 \left(
 \begin{array}{cc}
  \delta^{\alpha}_{\beta} & 0 \\
  0 & 0 \\
 \end{array}
 \right)
 \left(\begin{array}{c}
  E^{\beta}_J \\ F^{\beta}_J
 \end{array}\right) 
 =
 J(x)\,\widehat{\rm Re}\,J^{-1}(x),
\end{align}
whose complex representation is given by 
a map $\bbC^N\ni\tilde v\mapsto
v=J(x)\,\textrm{Re}\,[J^{-1}(x)\,\tilde v]\in\bbC^N$.%
\footnote{
  This expression first appeared in \cite{Fujii:2013sra} 
  as the projection to the tangent spaces to a Lefschetz thimble. 
} 
The other method does not require an explicit knowledge 
of the matrix elements of $J(x)$ 
(as in \cite{Alexandru:2017lqr}). 
We here demonstrate this procedure in the complex representation. 
We first look for two real column vectors $a,\,b\in\bbR^N$ 
such that they satisfy a linear equation
\begin{align}
 J(x)\,a+i\,J(x)\,b=\tilde v \in\bbC^N.
\label{linear_ab}
\end{align}
Here, $J(x)\,a$ and $J(x)\,b$ are obtained 
by numerically integrating \eqref{flow_zv_complex}
with the initial conditions 
$(z_0,v_0)=(x,a)$ and $(z_0,v_0)=(x,b)$, respectively.  
The linear equation \eqref{linear_ab} can be solved for $a,b$ 
with an iterative method. 
Once $a$ and $b$ are obtained, 
$v$ is given by $J(x)\,a$ 
(which we already have in the above process) 
because the formal solution for $a$ 
is given by $a=\textrm{Re}\,[J^{-1}(x)\,\tilde{v}]$.

We summarize the algorithm 
for a molecular dynamics on $T^\ast\Sigma_t$ 
that updates a configuration 
from $\zeta=({z},\pi)\in T^\ast\Sigma_t$ with ${z}={z}_t(x)$ 
to $\zeta'=({z}',\pi')=\Phi_{\Delta s}({z},\pi)
\in T^\ast\Sigma_t$ with ${z}'={z}_t(x')$. 
Every step below will be mostly given in the real representation, 
which can be readily translated to the complex representation.%
\footnote{
  When $M_{IJ}=\sigma^2 \delta_{IJ}$, 
  eqs.~\eqref{newRATTLE1}--\eqref{newRATTLE5} 
  have the following complex representations 
  for $z=(z^i=x^i+i\,y^i)$ and $\pi=(\pi_i=\pi_{i,x}+i\,\pi_{i,y})$ 
  with $\partial S(z)\equiv (\partial_i S(z))$ 
  and $J=(\partial z^i/\partial x^\alpha)$:
  \begin{align}
    \tilde z&=z+(\Delta s/\sigma^2)\,\pi
    -(\Delta s^2/2\sigma^2)\,[\partial S(z)]^\ast, 
    \nonumber
    \\
    z'&=\tilde z - i J(z) \lambda\quad(\lambda\in\bbR^N),
    \nonumber
    \\
    \pi_{1/2}&=(\sigma^2/\Delta s)\,(z'-z),
    \nonumber
    \\
    \tilde\pi'&=\pi_{1/2}-(\Delta s/2)\,[\partial S(z')]^\ast,
    \nonumber
    \\
    \pi'&=J(z')\,{\rm Re}\,(J^{-1}(z')\,\tilde\pi').
    \nonumber
  \end{align}
} 

\begin{description}

\item[Step 1.]
For $\zeta=({z},\pi)\in T^\ast \Sigma_t$ with ${z}={z}_t(x)$, 
we find $w=(u,\lambda)^T=((u^{\alpha}),\,(\lambda^{\alpha}))^T$ $({\alpha}=1,\ldots,N)$ 
that satisfies \eqref{ul_eq}. 
The equation can be solved iteratively, $w_k\to w_{k+1}=w_k + \Delta w$,
with Newton's method, 
starting from an initial guess $w_0=(u_0,\lambda_0)^T$ 
and solving the linear equation \eqref{recursion2} 
to obtain $\Delta w=(\Delta u,\Delta \lambda)^T$. 
Equation \eqref{recursion2} can be solved 
with either of a direct method or an iterative method. 
After $w=(u,\lambda)^T$ is obtained, 
we set $x'\equiv x+u$ and ${z}' \equiv {z}_t(x')$. 

\item[Step 2.]
Define
\begin{align}
 \pi_{1/2} \equiv \frac{1}{\Delta s}\,M ({z}'-{z}),
\label{newRATTLE3a}
\end{align}
and set 
\begin{align}
 \tilde{\pi}' \equiv \pi_{1/2} - \frac{\Delta s}{2}\, \partial V({z}').
\label{newRATTLE4a}
\end{align}

\item[Step 3.]
Project $\tilde{v}'\equiv M^{-1} \tilde\pi'$  
to $v'=\mathcal{P}({z}')\,\tilde{v}'\in T_{{z}'}\Sigma_t$  
to obtain 
$\pi'=M v'\in T^\ast_{{z}'}\Sigma_t$. 
The projection can be made with a direct method  
when the matrix elements of $J(x')$ are known explicitly, 
or with an iterative method. 
\end{description}

In practice, 
since $\Delta s$ is finite, 
it can happen, 
for ${z}$ close to zeros of the weight $e^{-S(z)}$, 
that 
one cannot find a solution $z'=z_t(x')$ to \eqref{Newton} 
anywhere in $\bbC^N$ 
or can only find a solution in a region beyond the zeros 
(see appendix \ref{sec:more_on_zeros} for detailed discussions). 
When this happens, 
we replace $\Phi_{\Delta s}$ by a {\em momentum flip} $\Psi$ 
that is defined by%
\footnote{
 Note that such an aggressive withdrawal is allowed as an algorithm 
 because a detour to go around zeros are provided by the tempering. 
} 
\begin{align}
  \Psi({z},\pi) = ({z},-\pi).
\end{align}

\begin{figure}[ht]
\centering
\includegraphics[width=55mm]{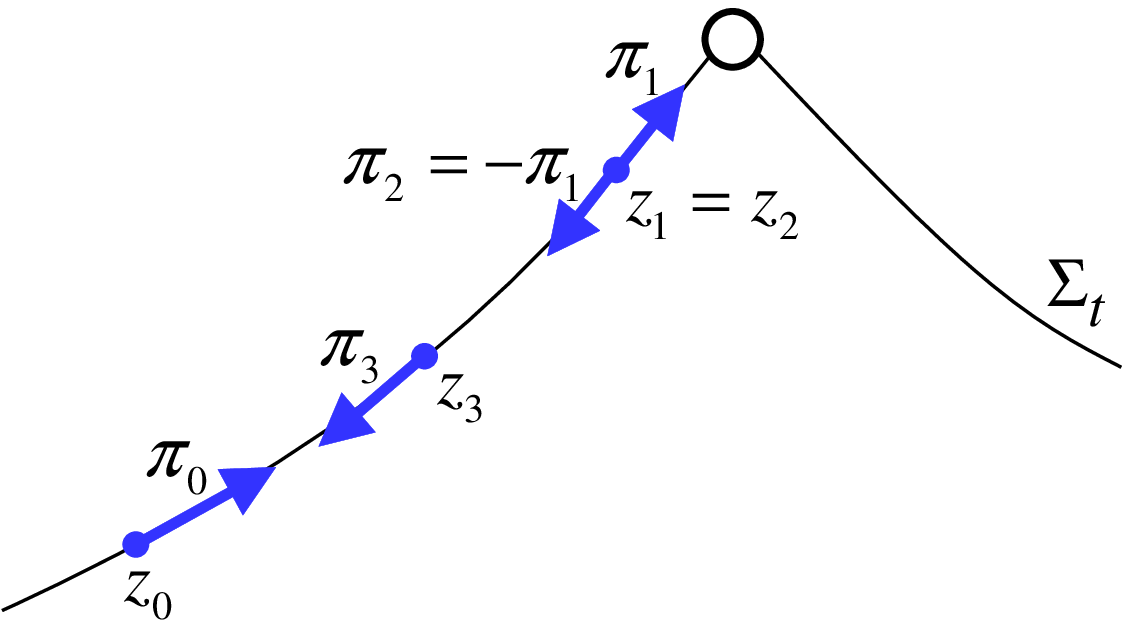} \hspace{5mm}
\includegraphics[width=55mm]{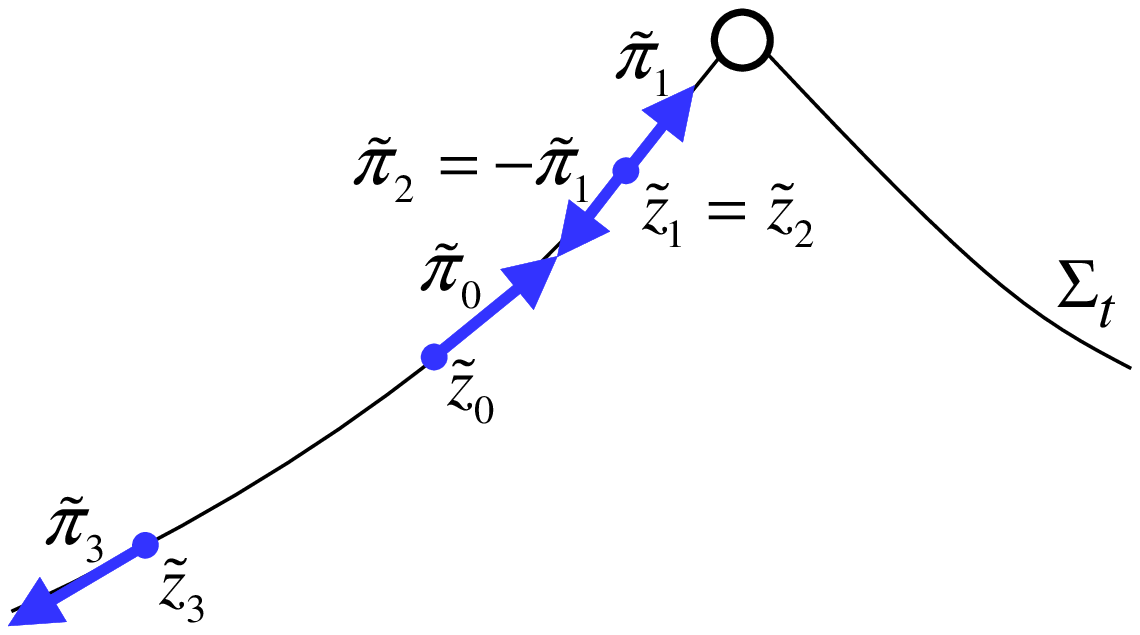} 
\caption{
\label{fig:flip}
 Moves on $\Sigma_t$ near a zero of $e^{-S(z)}$ (indicated by a circle). 
 (Left) An original forward move 
 with a momentum flip at ${z}_1$. 
 (Right) The time-reversed backward move. 
}
\end{figure}

Note that $\Psi$ is also volume-preserving 
[i.e., $(d{z}')_\parallel (d\pi')_\parallel
=(d{z})_\parallel(d\pi)_\parallel$ for $(z',\pi')=\Psi(z,\pi)$]
and reversible 
[i.e., if $\Psi({z},\pi) = ({z}',\pi')$ then 
$\Psi({z}',-\pi') = ({z},-\pi)$]. 
To understand the reversibility, 
consider a move of a molecule 
in the forward and backward directions in $s$, 
each consisting of three steps (see Fig.~\ref{fig:flip}): 
\begin{align}
 \mbox{forward:} &\quad
 ({z}_0,\pi_0)\xrightarrow{\Phi_{\Delta s}}({z}_1,\pi_1)
 \xrightarrow{\Psi}({z}_2,\pi_2)\xrightarrow{\Phi_{\Delta s}}({z}_3,\pi_3),
\\
 \mbox{backward:} &\quad
 (\tilde{z}_0,\tilde\pi_0)\xrightarrow{\Phi_{\Delta s}}
 (\tilde{z}_1,\tilde\pi_1)\xrightarrow{\Psi}
 (\tilde{z}_2,\tilde\pi_2)\xrightarrow{\Phi_{\Delta s}}
 (\tilde{z}_3,\tilde\pi_3)
\end{align}
with $(\tilde {z}_0,\tilde\pi_0)\equiv({z}_3,-\pi_3)$. 
\noindent
There, 
we assume that 
a move from ${z}_1$ with $\pi_1$ is prohibited 
by a prescribed condition. 
We thus flip the momentum from $\pi_1$ to $\pi_2=-\pi_1$, 
and the molecule arrives at ${z}_3$ with $\pi_3$. 
As for the backward move starting from $\tilde{z}_0={z}_3$
with $\tilde\pi_0=-\pi_3$, 
it will arrive at $\tilde{z}_1={z}_2\,(=z_1)$ 
with $\tilde\pi_1=-\pi_2\,(=\pi_1)$ 
thanks to the reversibility of $\Phi_{\Delta s}$. 
Then the further move with $\tilde\pi_1$ must be prohibited 
by the same condition 
that prohibited the further move from $z_1$ with $\pi_1$. 
Then we make a momentum flip from $\tilde\pi_1$ to $\tilde\pi_2=-\tilde\pi_1$, 
and the molecule will arrive at $\tilde{z}_3$ 
which must coincide with ${z}_0$ 
again thanks to the reversibility of $\Phi_{\Delta s}$.
We thus see that the reversibility holds for the whole process 
with the relation 
$(\tilde{z}_n,\tilde\pi_n)=({z}_{3-n},-\pi_{3-n})$.

Due to the volume-preservation and the reversibility 
of $\Phi_{\Delta s}$ and $\Psi$, 
the transition probability 
\begin{align}
 \mathbb{P}^{(1)}(\zeta'|\zeta)
 \equiv 
 \min\bigl(1,e^{-H(\zeta')+H(\zeta)}\bigr)\,
 \delta(\zeta'-\Phi_{\Delta s}^{\,n}(\zeta))
 \quad(\zeta'\neq\zeta)
\end{align}
satisfies the following relation [see \eqref{W(1)} and \eqref{W(1)_det}]:
\begin{align}
 \mathbb{P}^{(1)}(\zeta'|\zeta)\,e^{-H(\zeta)}
 = \mathbb{P}^{(1)}(\zeta^T|\zeta^{\prime\,T})\,e^{-H(\zeta')}
 \quad(\zeta'\neq\zeta)
\end{align}
even when the partial replacements from $\Phi_{\Delta s}$ to $\Psi$ 
are made. 
In the following, 
we only use the symbol $\Phi_{\Delta s}$ 
with the understanding that it will be replaced by $\Psi$ when necessary. 

\section{HMC on TLTM}
\label{sec:HMC_partial}

In this section, 
after introducing a method to swap configurations at adjacent replicas, 
we summarize the HMC algorithm on TLTM.

\subsection{Swap of configurations at adjacent replicas}
\label{sec:swap}

\begin{figure}[ht]
\centering
\includegraphics[width=80mm]{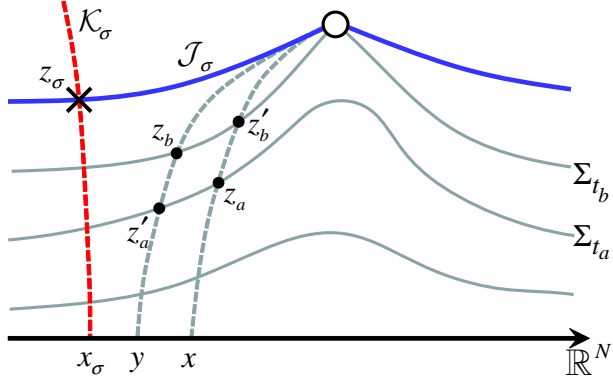}%
\caption{
\label{fig:swap}
 Swap of configurations 
 from $(z_a, z_b)\equiv (z_{t_a}(x),z_{t_b}(y))
 \in\Sigma_{t_a}\times\Sigma_{t_b}$ 
 to 
 $(z'_a, z'_b)\equiv (z_{t_a}(y),z_{t_b}(x))
 \in\Sigma_{t_a}\times\Sigma_{t_b}$,
 which is actually the exchange of the initial configurations $x$ and $y$.  
 In figure, 
 $\mathcal{J}_\sigma$ is the Lefschetz thimble 
 associated to a critical point ${z}_\sigma$. 
 $\mathcal{K}_\sigma$ is the corresponding anti-thimble.
 The distribution $\propto e^{-V(z)}=e^{-{\rm Re}\,S(z)}$ on $\Sigma_{t}$ 
 has peaks at intersection points of $\Sigma_t$ and $\mathcal{K}_\sigma$. 
}
\end{figure}

\begin{figure}[ht]
\centering
\includegraphics[width=80mm]{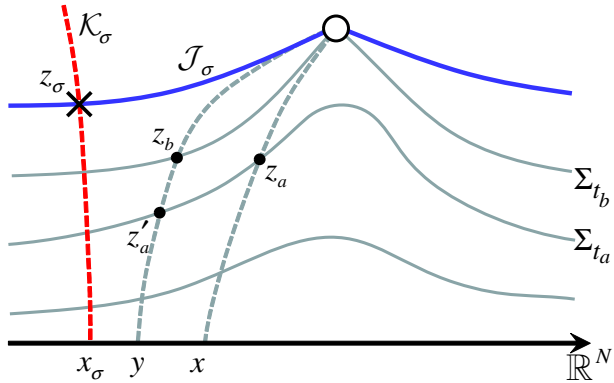}%
\caption{
\label{fig:no-swap}
  The case when $z_t(x)$ reaches a zero at $t<t_b$.
  We then simply reject the proposal for the swap $(x,y)\to(x',y')=(y,x)$. 
}
\end{figure}

We realize the swap of configurations at adjacent replicas, 
$\Sigma_{t_a}$ and $\Sigma_{t_b}$ ($b=a\pm1$), 
by the {\em exchange of the initial configurations}. 
Namely,
$(z_a, z_b)\equiv (z_{t_a}(x),z_{t_b}(y))
\in\Sigma_{t_a}\times\Sigma_{t_b}$ 
is proposed to be updated to 
$(z'_a, z'_b)\equiv (z_{t_a}(x'),z_{t_b}(y'))
\in\Sigma_{t_a}\times\Sigma_{t_b}$ 
with $(x',y')=(y,x)$ 
(see Fig.~\ref{fig:swap}).
\noindent
Accordingly, 
the accept/reject probability must be 
with respect to $(x,y)\in\bbR^N\times\bbR^N$, 
and the algorithm takes the following form:
\begin{enumerate}
\item
We first calculate the Jacobian matrices 
$J_a \equiv J_{t_a}(x)$, $J_b \equiv J_{t_b}(y)$. 

\item
We further calculate 
${z}_a'\equiv {z}_{t_a}(y)$ and ${z}_b'\equiv {z}_{t_b}(x)$ 
together with the corresponding Jacobian matrices, 
$J_a' \equiv J_{t_a}(y)$ and $J_b' \equiv J_{t_b}(x)$.

\item
We update the original initial configurations $(x,y)$ 
to the swapped initial configurations $(x',y')=(y,x)$
with the probability%
\footnote{
  For $z_a=z_{t_a}(x)$ and $z_b=z_{t_b}(y)$ with $t_a<t_b$, 
  $z_t(x)$ may reach a zero at $t<t_b$ 
  (see Fig.~\ref{fig:no-swap}). 
  In such a case,  
  we reject the proposal for the swap $(x,y)\to(x',y')=(y,x)$. 
  Note that this procedure keeps the relation \eqref{two-body}. 
\label{fn:no-swap}
} 
\begin{align}
 \min\left(1,
 \frac{ |\det J_a'|~|\det J_b'|\, 
 e^{-V({z}_a') - V({z}_b')}}
 { |\det J_a|~|\det J_b|\, 
 e^{-V({z}_a) - V({z}_b)}}
 \right). 
\label{swap_x}
\end{align}
\end{enumerate}
\noindent

The above procedure correctly leads to 
the global equilibrium 
on the product space 
$\Sigma_{\rm tot}=\Sigma_{t_0}\times\cdots\times\Sigma_{t_A}$  
with the distribution 
$ p_\mathrm{eq}(\vec{z})\,\prod_{a=0}^A|dz_a|
 \propto\prod_{a=0}^A e^{-V(z_a)}\, |dz_a|$. 
In fact, 
by using the identity $|d{z}|=|\det J(x)|\,dx$, 
the transition probability $\hat{P}^{(2)}_{ab}(x',y'|x,y)$ 
from $(x,y)\in\bbR^{N}\times\bbR^N$ to $(x',y')\in\bbR^{N}\times\bbR^N$ 
can be rewritten to the transition probability 
$P^{(2)}_{ab}(z_a',z_b'|z_a,z_b)$  from 
$(z_a, z_b)= (z_{t_a}(x),z_{t_b}(y))
\in\Sigma_{t_a}\times\Sigma_{t_b}$ 
to 
$(z'_a, z'_b)= (z_{t_a}(x'),z_{t_b}(y'))
\in\Sigma_{t_a}\times\Sigma_{t_b}$ 
as follows:
\begin{align}
 P^{(2)}_{ab}(z_a',z_b'|z_a,z_b)
 &=|\det J_{t_a}(x')|^{-1}\,|\det J_{t_b}(y')|^{-1}\,
 \hat{P}^{(2)}_{ab}(x',y'|x,y),
\end{align}
which means that the transition probability for $(x,y)\to (x',y')$ 
is translated to the following probability 
for $({z}_a,{z}_b)\to({z}'_a,{z}'_b)$:
\begin{align}
 P^{(2)}_{ab}(z_a',z_b'|z_a,z_b)
 &=\min\left( \frac{1}{|\det J'_a|\,|\det J'_b|},\,
 \frac{ e^{-V({z}_a') - V({z}_b')}}
 { |\det J_a|~|\det J_b|\, 
 e^{-V({z}_a) - V({z}_b)}}
 \right)
\nonumber
\\
 &~~~\times \delta(x'-y)\,\delta(y'-x) 
 \quad [(z_a',z_b')\neq(z_a,z_b)]. 
\end{align}
Then one can easily show that 
the following detailed balance condition does hold:
\begin{align}
 P^{(2)}_{ab}(z_a',z_b'|z_a,z_b)\,e^{-V(z_a)-V(z_b)}
 = P^{(2)}_{ab}(z_a,z_b|z_a',z_b')\,e^{-V(z_a')-V(z_b')}.
\end{align}
Note that 
we have to calculate the Jacobian determinant explicitly 
at every swapping process 
even though this is not mandatory for the molecular dynamics 
on each flowed surface. 

We make an important comment on the reason 
why we use initial configurations as a reference 
in the swapping process. 
In general, 
one can introduce an arbitrary coordinate system to each flowed surface, 
to be used as a reference in swapping configurations as above. 
However, 
for such arbitrarily chosen coordinate systems, 
the distributions as functions of the coordinates 
will take very different functional forms between adjacent replicas, 
and one cannot expect a significant acceptance rate. 
On the other hand, 
this problem will not occur 
if we take the initial configurations as a common reference, 
because the distributions then have peaks at the same coordinate values 
(such as $x_\sigma$ in Fig.~\ref{fig:swap} 
that flows to a critical point ${z}_\sigma$) 
for different flowed surfaces \cite{Fukuma:2019wbv}. 

Another comment is that 
one can extend the molecular dynamics 
to the phase space of the whole enlarged configuration space, 
$T^\ast\Sigma_{t_0}\times\cdots\times T^\ast\Sigma_{t_A}$, 
also by swapping momenta $\pi_a$ 
in the course of molecular dynamics, 
as in \cite{Sugita1999}. 
However, the additional computational cost will not be negligible, 
because in TLTM 
we need to transport $\pi_a\in T^\ast\Sigma_{t_a}$ 
({\em resp.} $\pi_b\in T^\ast\Sigma_{t_b}$)
to obtain $\pi'_b\in T^\ast\Sigma_{t_b}$ 
({\em resp.} $\pi'_a\in T^\ast\Sigma_{t_a}$), 
and such a transport generically causes 
an additional difference in the sum of Hamiltonians, 
which lowers acceptance rates. 
We leave the investigation of this algorithm for future work. 

\subsection{Summary of HMC on TLTM}
\label{sec:HMC_summary}

We summarize the HMC algorithm on the TLTM 
by following the outline given in subsection \ref{sec:outline} 
(recall that $V(z) = {\rm Re}\,S(z)$):

\begin{description}

\item[Step A. HMC on $\{\Sigma_{t_a}\}$:]
~

\begin{description}

\item[Step A1 (initial setup).]
For a given configuration $\vec{z}=({z}_a={z}_{t_a}(x_a))
\in \Sigma_{\rm tot}=\Sigma_{t_0}\times\cdots\times\Sigma_{t_A}$, 
we generate $\vec{\tilde{\pi}}=(\tilde\pi_{a})
\in T_{{z}_0}^\ast\bbR^{2N}\times\cdots\times T_{{z}_A}^\ast\bbR^{2N}$ 
with a Gaussian distribution 
$\propto \prod_a e^{-\tilde\pi_{a}^2/2M_a}$, 
and for every $a$ 
we project $\tilde\pi_a\in T_{{z}_a}^\ast\bbR^{2N}$ 
to $\pi_a=\tilde\pi_{a}\,\mathcal{P}({z}_a)\in T^\ast_{{z}_a}\Sigma_{t_a}$. 
The projection can be made with a direct method  
when the matrix elements of $J_a\equiv J_{t_a}(x_a)$ are known explicitly, 
or with an iterative method. 

\item[Step A2.] 
For each $\zeta_a=({z}_a,\pi_a)\in T^\ast \Sigma_{t_a}$ 
with ${z}_a={z}_{t_a}(x_a)$, 
we find $w=(u,\lambda)^T$ 
that satisfies \eqref{ul_eq}. 
The equation can be solved iteratively, $w_k\to w_{k+1}=w_k + \Delta w$,
with Newton's method, 
starting from an initial guess $w_0=(u_0,\lambda_0)^T$ 
and solving the linear equation \eqref{recursion2} 
to obtain $\Delta w=(\Delta u,\Delta \lambda)^T$. 
Equation \eqref{recursion2} can be solved 
with either of a direct method or an iterative method. 
After $w=(u,\lambda)^T$ is obtained, 
we set $x'_a = x_a+u$ and ${z}'_a = {z}_{t_a}(x'_a)$. 

\item[Step A3.]
For each replica $a$, 
we define
\begin{align}
 \pi_{a,1/2} \equiv \frac{1}{\Delta s}\,M ({z}'_a-{z}_a),
\label{newRATTLE3a_v2}
\end{align}
and set 
\begin{align}
 \tilde{\pi}'_a \equiv \pi_{a,1/2} - \frac{\Delta s}{2}\, 
 \partial\, {\rm Re}S({z}'_a).
\label{newRATTLE4a_v2}
\end{align}

\item[Step A4.]
For each replica $a$, 
we project $\tilde\pi'_a\in T_{{z}'_a}^\ast\bbR^{2N}$ 
to $\pi'_a=\tilde\pi'_{a}\,\mathcal{P}({z}'_a)
\in T^\ast_{{z}'_a}\Sigma_{t_a}$. 
The projection can be made with a direct method  
when the matrix elements of $J'_a \equiv J_{t_a}(x'_a)$ 
are known explicitly, 
or with an iterative method. 

\item[Step A5.]
We repeat Steps A2 through A4 a fixed number of times $(\equiv n)$ 
for all replicas.

\item[Step A6.]
For each replica $a$, 
we calculate $\Delta H_a \equiv H_a(\zeta'_a)-H_a(\zeta_a)$, 
and update $\zeta_a$ to $\zeta'_a$
with the probability $\min(1,e^{-\Delta H_a})$.

\item[Step A7.]
We ignore the values of $\pi'_a$ and only keep those of $z'_a$.

\end{description}

\item[Step B. Swap among $\{\Sigma_{t_a}\}$:]
~

\begin{description}
\item[Step B1.]
For a given pair 
$({z}_a,{z}_b)=({z}_{t_a}(x),{z}_{t_b}(y))\in\Sigma_{t_a}\times\Sigma_{t_b}$, 
we calculate the Jacobian matrices 
$J_a \equiv J_{t_a}(x)$, $J_b \equiv J_{t_b}(y)$ 
by numerically integrating the flow equations 
\eqref{flow_z} and \eqref{flow_J}
if they have not been calculated yet. 

\item[Step B2.]
We calculate 
${z}_a'\equiv {z}_{t_a}(y)$ and ${z}_b'\equiv {z}_{t_b}(x)$ 
together with the corresponding Jacobian matrices, 
$J_a' \equiv J_{t_a}(y)$ and $J_b' \equiv J_{t_b}(x)$. 

\item[Step B3.]
We update the original initial configurations $(x,y)$ 
to the swapped initial configurations $(x',y')=(y,x)$
with the probability%
\footnote{
  See a notice in footnote \ref{fn:no-swap}.
} 
\begin{align}
 \min\left(1,
 \frac{ |\det J_a'|~|\det J_b'|\, 
 e^{-{\rm Re}\,S({z}_a') - {\rm Re}\,S({z}_b')}}
 { |\det J_a|~|\det J_b|\, 
 e^{-{\rm Re}\,S({z}_a) - {\rm Re}\,S({z}_b)}}
 \right).
\end{align}
\end{description}

\item[Step C. Measurement:]
~

After repeating Steps A and B sufficiently many times, 
we make a measurement 
and save the values $\{e^{i\theta({z}_a)},\,\mathcal{O}({z}_a)\}$ 
$(a=0,\ldots,A)$, 
that are calculated from ${z}_a$ and $J_a$.
\end{description}
%

\section{Results and analysis}
\label{sec:results_analysis}

In this section, we apply 
the TLTM to the Hubbard model 
both with  HMC (implemented in this paper) 
and with Metropolis (adopted in \cite{Fukuma:2017fjq,Fukuma:2019wbv}). 
We first confirm both algorithms to work properly 
by showing that 
the expectation value of the number density operator is estimated correctly. 
We then show that HMC is more efficient than Metropolis  
even for small degrees of freedom ($N=20$). 

\subsection{Hubbard model and the parameters for simulations}
\label{sec:settings}

Let $\Lambda$ be a lattice with $N_s$ sites. 
The Hubbard model on $\Lambda$ is defined by the Hamiltonian 
\begin{align}
 H &= {}  -\kappa\,\sum_{\bx,\by}\sum_\sigma\,K_{\bx \by}\,
 c_{\bx,\sigma}^\dagger c_{\by,\sigma}
 - \mu\,\sum_\bx\,(n_{\bx,\uparrow} + n_{\bx,\downarrow} - 1)
\nonumber
\\
 & ~~~~ + U \sum_\bx\,(n_{\bx,\uparrow}-1/2)\,(n_{\bx,\downarrow}-1/2). 
\end{align}
Here, $c_{\bx,\sigma}$ and $c_{\bx,\sigma}^\dagger$ 
are the annihilation and creation operators 
on site $\bx\in\Lambda$ with spin $\sigma$ $(=\uparrow,\downarrow)$, 
obeying 
$\{c_{\bx,\sigma},c_{\by,\tau}^\dagger\} 
= \delta_{\bx \by}\,\delta_{\sigma \tau}$ and 
$\{c_{\bx,\sigma},c_{\by,\tau}\}
= \{c_{\bx,\sigma}^\dagger,c_{\by,\tau}^\dagger\} = 0$, 
and $n_{\bx,\sigma}\equiv c_{\bx,\sigma}^\dagger c_{\bx,\sigma}$. 
$K_{\bx \by}$ is the adjacency matrix 
that takes a nonvanishing value ($\equiv 1$) only for nearest neighbors, 
and we assume the lattice to be bipartite. 
$\kappa\,(\,>0)$ is the hopping parameter, $\mu$ is the chemical potential, 
and $U\,(\,>0)$ represents the strength of the on-site repulsive potential. 
We have shifted $n_{\bx,\sigma}$ as $n_{\bx,\sigma}-1/2$
so that $\mu=0$ corresponds to the half-filling state, 
$\sum_\sigma\langle n_{\bx,\sigma}-1/2 \rangle = 0$. 

By using the Trotter decomposition with equal spacing $\epsilon$ 
($\beta = N_\tau \epsilon$), 
we can rewrite the expectation value of the number density 
$n\equiv (1/N_s)\sum_{\bx}(n_{\bx,\uparrow}+n_{\bx,\downarrow}-1)$ 
as a path integral over a Gaussian Hubbard-Stratonovich variable 
$\phi=(\phi_{\ell,\bx})$ as follows 
(see, e.g., \cite{Fukuma:2019wbv} for the derivation):
\begin{align}
  \langle n \rangle 
  &\equiv \frac{\int d\phi\,e^{-S(\phi)}\,n(\phi)}
    {\int d\phi\,e^{-S(\phi)}} 
    \quad
    \Bigl( d\phi \equiv \prod_{\ell,\bx} d\phi_{\ell,\bx} \Bigr),
    \label{n_PI}
\\
  e^{-S(\phi)} &\equiv e^{-(1/2)\,\sum_{\ell,\bx} \phi_{\ell,\bx}^2}\,
                 \det M^a(\phi)\,\det M^b(\phi),
\label{Hubbard_action1}
\\
  M^{a/b}(\phi) 
  &\equiv \pmb{1} + e^{\pm \beta \mu}\,\prod_\ell e^{\epsilon \kappa K}\, 
    e^{\pm \,i\sqrt{\epsilon U} \phi_\ell},
    \label{Hubbard_action2}
\\
  n(\phi) & \equiv (i\sqrt{\epsilon U} N_\tau N_s)^{-1}\,
  \sum_{\ell,\bx} \phi_{\ell,\bx}, 
\label{Hubbard_action3}
\end{align}
where $\phi_\ell\equiv(\phi_{\ell,\bx}\,\delta_{\bx\by})$ 
and $\prod_\ell$ is a product in descending order. 
Below we apply the TLTM to this model, 
in which the variables $\phi=(\phi_{l,\bx})$ 
correspond to $x=(x^i)$  
with $i=1,\ldots,N \,(\equiv N_\tau N_s)$. 

We use a two-dimensional periodic square lattice 
of size $2\times 2$ (thus $N_s=4$). 
The parameters in the Hamiltonian 
are set to $\beta\kappa=3$, $\beta\mu=4$, $\beta U = 13$. 
The imaginary time is decomposed to $N_\tau=5$ pieces. 
We set $A=6$, $T=0.24$ (maximum flow time), 
and $t_a$'s are set linearly in $a$.%
\footnote{
  See \cite{Fukuma:2019wbv} for a justification of the linear spacing 
  that is based on the geometrical optimization 
  \cite{Fukuma:2017wzs,Fukuma:2018qgv}.
} 
We use as an initial configuration the one obtained after a test run. 
The same initial configuration is used for HMC and Metropolis . 
After discarding $2,000$ configurations to ensure equilibration, 
we take ${N_{\rm conf}} = 30,000$ configurations for estimations. 
In both algorithms, 
we first perform swapping process, 
then make transitions on each flowed surface, 
and finally make measurements. 

In molecular dynamics, 
we set the step size to $\Delta s = 0.1$ 
and the step number to $n=10$. 
We set $\sigma_a^2 = 1$ in $(M_a)_{IJ}=\sigma_a^2\,\delta_{IJ}$ for all $a$, 
and use the direct methods for the inversions 
in Steps A1, A2 and A4 in subsection~\ref{sec:HMC_summary}.%
\footnote{
  The direct method (LU decomposition) turns out to be faster 
  than the iterative method (BiCGStab) 
  for the case we consider with small degrees of freedom.
} 
In solving \eqref{recursion2} iteratively, 
we set the initial guess to $w_0 = 0$, 
and rescale $\Delta w$ as $\Delta w \rightarrow 0.15 \times \Delta w$ 
when $|\Delta w|/\sqrt{2N}$ is
larger than $0.5\times\Delta s/|\det J_{t_a}(x)|^{1/N}$ 
or when $|f(w_k+\Delta w)| > |f(w_k)|$ 
(see appendix \ref{sec:more_on_zeros}
for a detailed explanation on the conditions for the scaling). 
We set the stopping criterion  
to $|\Delta w|/\sqrt{2N}<10^{-8}$ 
{\it and} $|f(w_k)|/\sqrt{2N}<10^{-5}$. 
We find that the process terminates with $3-7$ iterations in most cases. 
The momentum is flipped either when the number of iterations exceeds 50 
or when $x+u_k+\Delta u$ still flows to a zero of $e^{-S(z)}$ 
even after the rescaling $\Delta w \rightarrow 0.15^3 \times \Delta w$ 
(see also appendix \ref{sec:more_on_zeros} 
on the condition for the replacement). 
We have tested the reversibility for some configurations 
chosen in the vicinity of zeros of $e^{-S(z)}$, 
and found that $|z'-z|/\sqrt{2N}$ is around $O(10^{-10})$  
for $(z',\pi') 
= (\Phi_{\Delta s}^{n=10}\circ\Psi\circ\Phi_{\Delta s}^{10})({z},\pi)$ 
where $\pi$ is generated from a Gaussian distribution with unit variance. 

For Metropolis, 
we use the isotropic Gaussian proposals 
of the standard deviations $\sigma_\mathrm{Met} = 0.037 - 0.18$ 
(varying on replicas), 
which are tuned 
so that the acceptance rate is 0.5 -- 0.7 (see Table \ref{table:hmc_met_param}).  
We repeat this procedure $n_\mathrm{Met}=50$ times. 

The swapping process is performed by pairing seven replicas 
in two different ways. 
One is (i): $(a,b)=(0,1),(2,3),(4,5)$ leaving replica 6 intact, 
and the other is (ii): $(1,2),(3,4),(5,6)$ leaving replica 0 intact. 
We repeat the swap seven times changing the pairing (i) and (ii) alternately.%
\footnote{
  By making independent pairs of replicas in this way, 
  the swap can be performed in parallel processes. 
} 
The acceptance rates thus obtained 
are shown in Table \ref{table:acceptance_rates}.%
\footnote{
  The acceptance rates for HMC may seem too large. 
  They can actually be reduced by increasing $\Delta s$, 
  but this must be done carefully, 
  because for too large $\Delta s$ 
  Newton's method in solving \eqref{Newton} may converge to unwanted solutions, 
  which violates the reversibility 
  (see discussions in appendix \ref{sec:more_on_zeros}). 
} 
\begin{table}[htb]
  \centering
  \begin{tabular}{|c|c|c|c|c|c|c|c|c|}
    \hline
    algorithm & parameter & $a=0$ & $1$ & $2$ & $3$ & $4$ & $5$ & $6$ \\
    \hline
    HMC & $\Delta s$ & 0.1 & 0.1 & 0.1 & 0.1 & 0.1 & 0.1 & 0.1 \\
              & $n$ & 10 & 10 & 10 & 10 & 10 & 10 & 10 \\
    \hline
    Metropolis & $\sigma_\mathrm{Met}$ & 0.18 & 0.13 & 0.10 & 0.081 & 0.066 & 0.038 & 0.037 \\
              & $n_\mathrm{Met}$ & 50 & 50 & 50 & 50 & 50 & 50 & 50 \\
    \hline
  \end{tabular}
  \caption{The parameters in HMC and Metropolis.}
  \label{table:hmc_met_param}
\end{table}\noindent
\begin{table}[htb]
  \centering
  \begin{tabular}{|c|c|c|c|c|c|c|c|c|}
    \hline
    algorithm & direction & $a=0$ & $1$ & $2$ & $3$ & $4$ & $5$ & $6$ \\
    \hline
    HMC (w/ swap) & $x$ (or $z$) & 0.99 & 0.98 & 0.96 & 0.95 & 0.94 & 0.94 & 0.94 \\
              & $a$ & 0.52 & 0.48 & 0.47 & 0.46 & 0.46 & 0.45 & - \\
    \hline
    Metropolis (w/ swap) & $x$ (or $z$) & 0.53 & 0.54 & 0.55 & 0.55 & 0.55 & 0.66 & 0.61 \\
              & $a$ & 0.52 & 0.48 & 0.47 & 0.46 & 0.46 & 0.45 & - \\
    \hline
    HMC (w/o swap)
              & $x$ (or $z$) & 0.99 & 0.98 & 0.96 & 0.95 & 0.94 & 0.91 & 0.96 \\
    \hline
    Metropolis (w/o swap) & $x$ (or $z$) & 0.53 & 0.54 & 0.55 & 0.55 & 0.55 & 0.57 & 0.67 \\
    \hline
  \end{tabular}
  \caption{The average acceptance rates.}
  \label{table:acceptance_rates}
\end{table}\noindent

Calculations without swap are also carried out for comparison, 
with the same parameters and the same initial configuration 
as those for the calculations with swap.

\subsection{Estimate of the number density}
\label{sec:number_density}

\begin{figure}[ht]
  \centering
  \includegraphics[width=60mm]{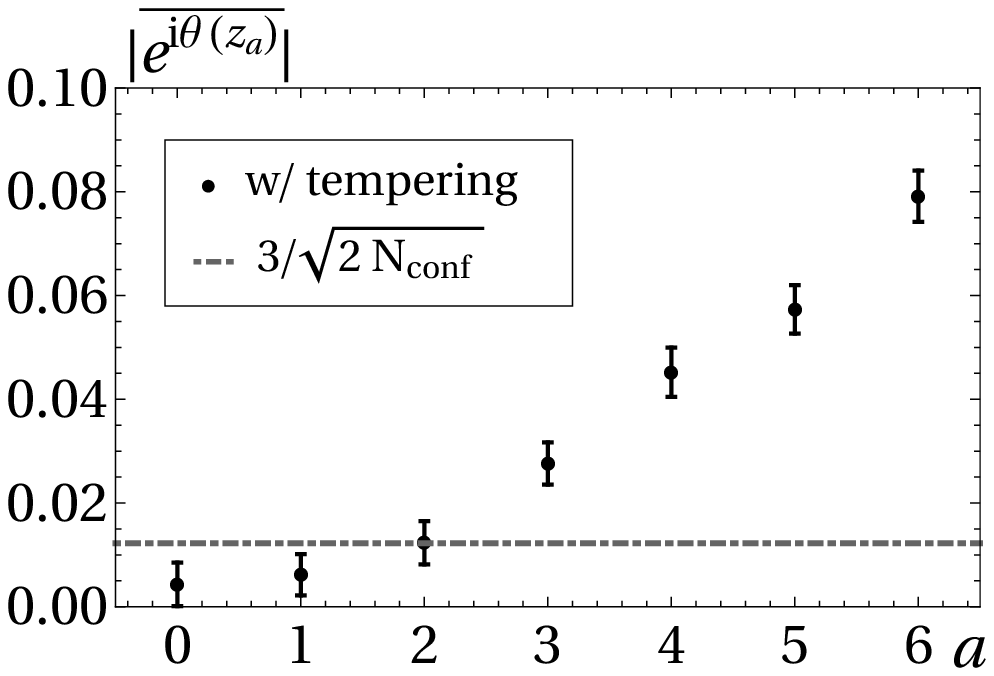}%
  \hspace{1cm}
  \includegraphics[width=60mm]{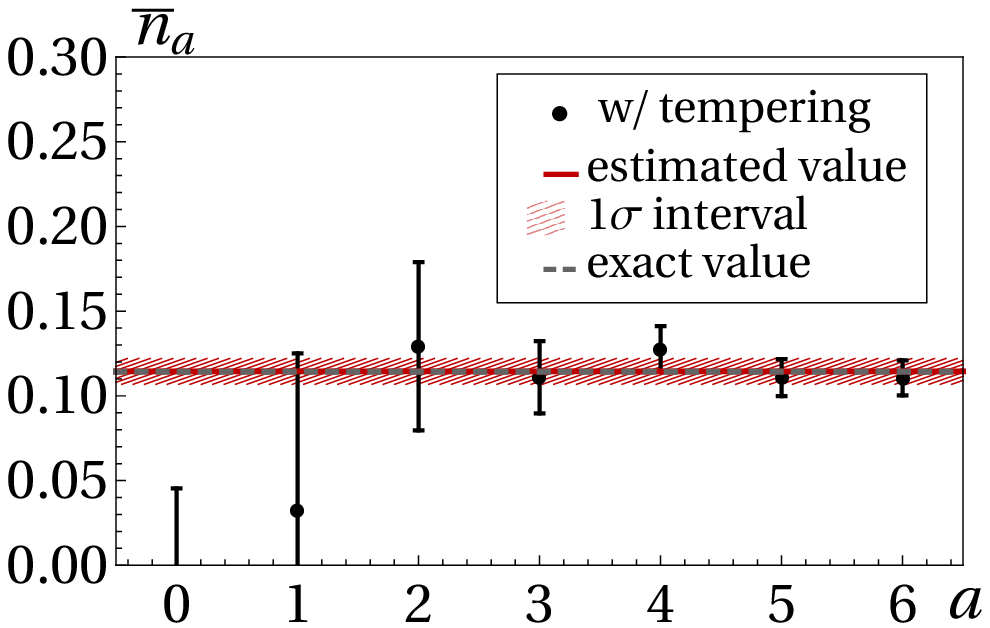} 
  \caption{
    \label{fig:hmc_results}
    The results for HMC with swap. 
    (Left) The sign averages. 
    (Right) The estimates $\bar{n}_a$. 
  }
\end{figure}

\begin{figure}[ht]
  \centering
  \includegraphics[width=60mm]{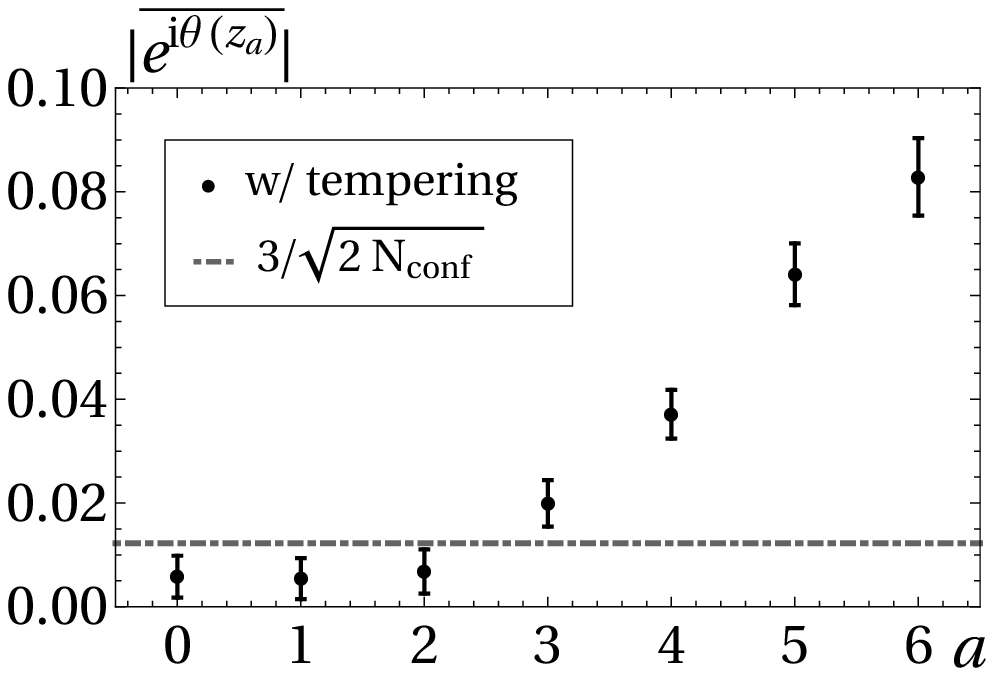}%
  \hspace{1cm}
  \includegraphics[width=60mm]{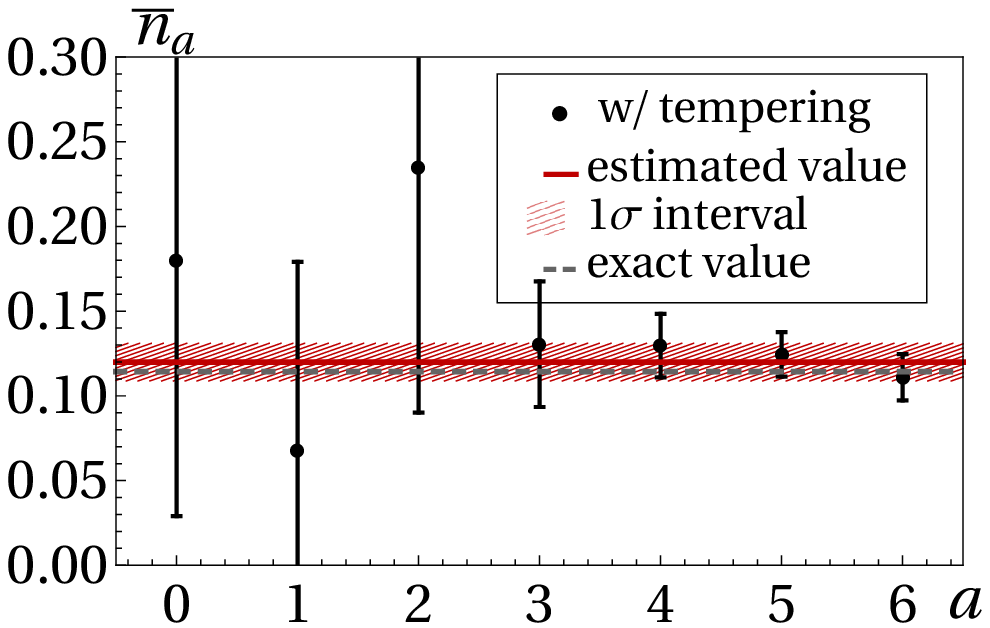} 
  \caption{
    \label{fig:met_results}
    The results for Metropolis with swap. 
    (Left) The sign averages. 
    (Right) The estimates $\bar{n}_a$. 
  }
\end{figure}

In order to confirm the algorithms to work correctly, 
we evaluate the expectation values of the number density, 
$\langle n \rangle$. 
The sign averages and the expectation values 
are shown in Figs.~\ref{fig:hmc_results} and \ref{fig:met_results}. 
\noindent
\noindent
We fit the data points in the range $a=3,\ldots,6$ 
for both HMC and Metropolis, 
which are chosen by observing 
that $\bigl|\overline{e^{i\theta(z_a)}}\bigr|$ 
are above $3/\sqrt{2N_{\rm conf}}$ including the error margin 
\cite{Fukuma:2019wbv}. 
The results are 
$\langle n \rangle \approx 0.1145 \pm 0.0076$ 
for HMC ($\chi^2/{\rm DOF} = 0.48$), 
and $\langle n \rangle \approx 0.120 \pm 0.011$ 
for Metropolis ($\chi^2/{\rm DOF} = 0.44$), 
that should be compared with 
the exact value $\langle n\rangle = 0.1143$ 
(the value under the Trotter decomposition \cite{Fukuma:2019wbv}), 
and thus we confirm that the algorithms work correctly. 

\begin{figure}[ht]
  \centering
  \includegraphics[width=60mm]{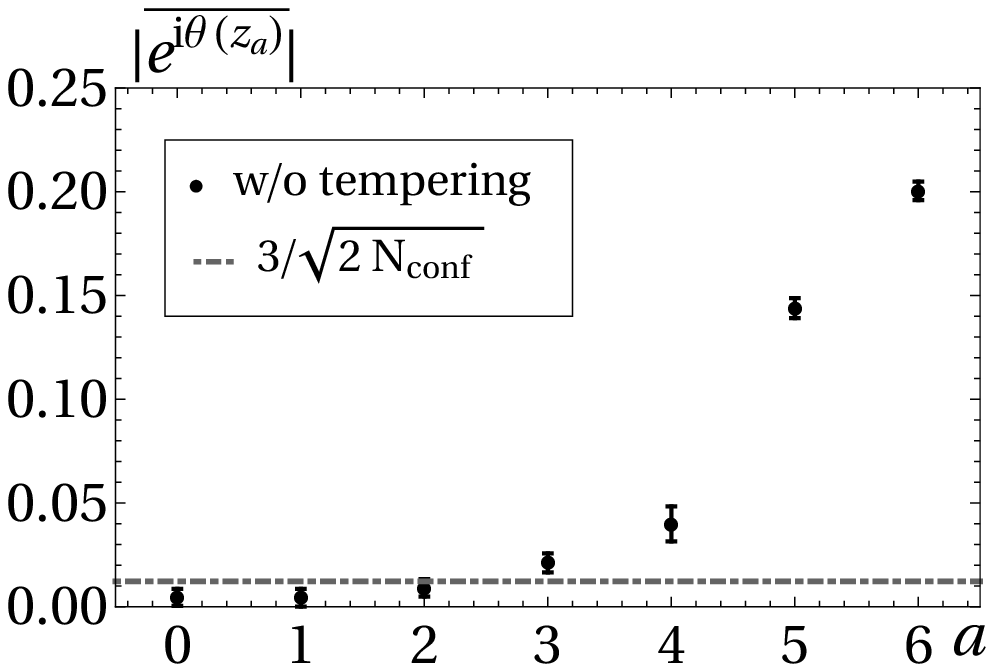}%
  \hspace{1cm}
  \includegraphics[width=60mm]{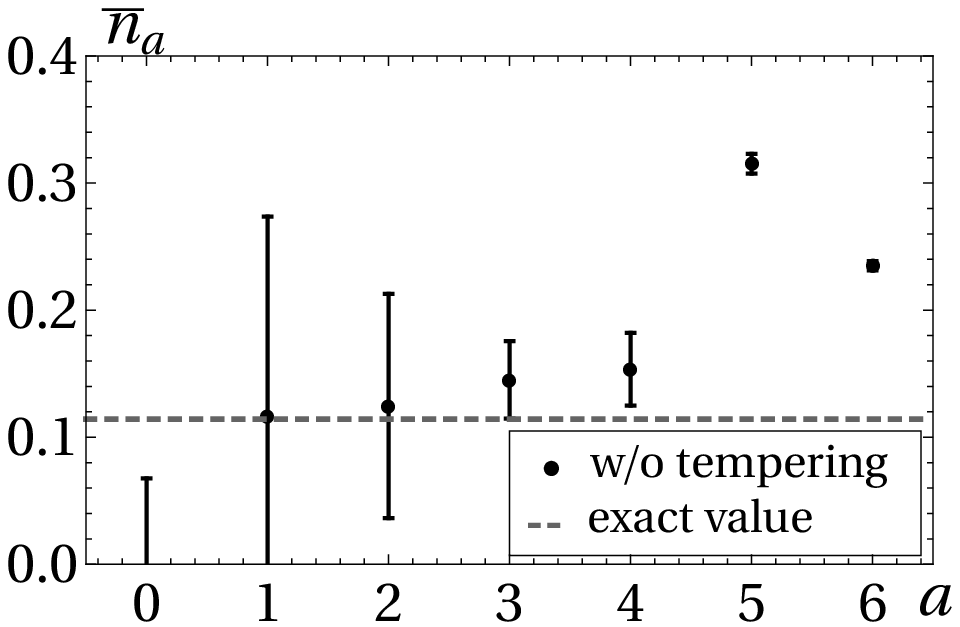} 
  \caption{
    \label{fig:hmc_results_noswap}
    The results for HMC without swap. 
    (Left) The sign averages. 
    (Right) The estimates $\bar{n}_a$. 
  }
\end{figure}

\begin{figure}[ht]
  \centering
  \includegraphics[width=60mm]{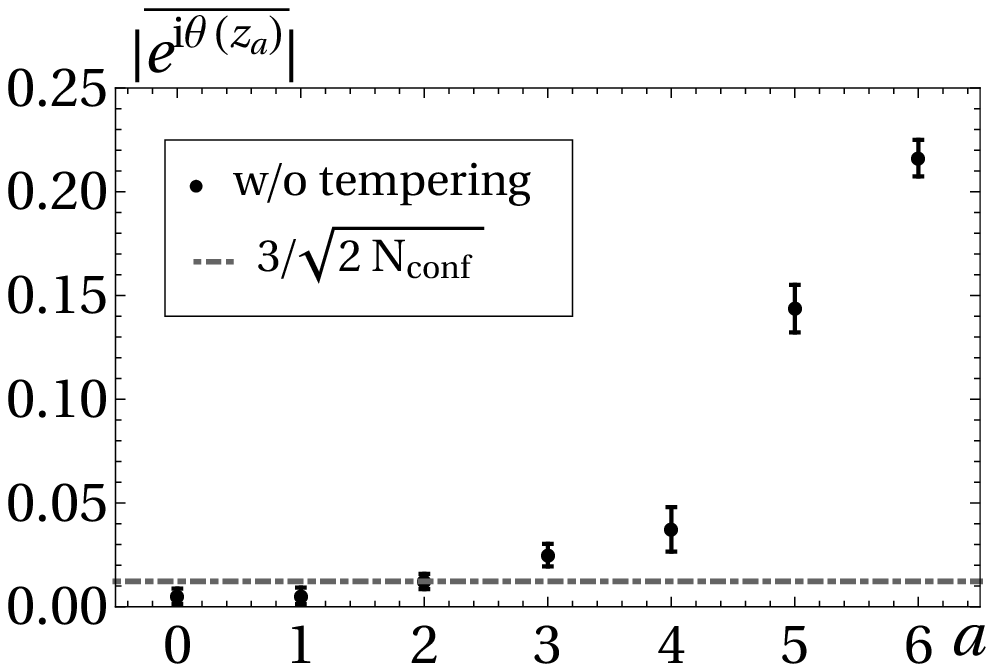}%
  \hspace{1cm}
  \includegraphics[width=60mm]{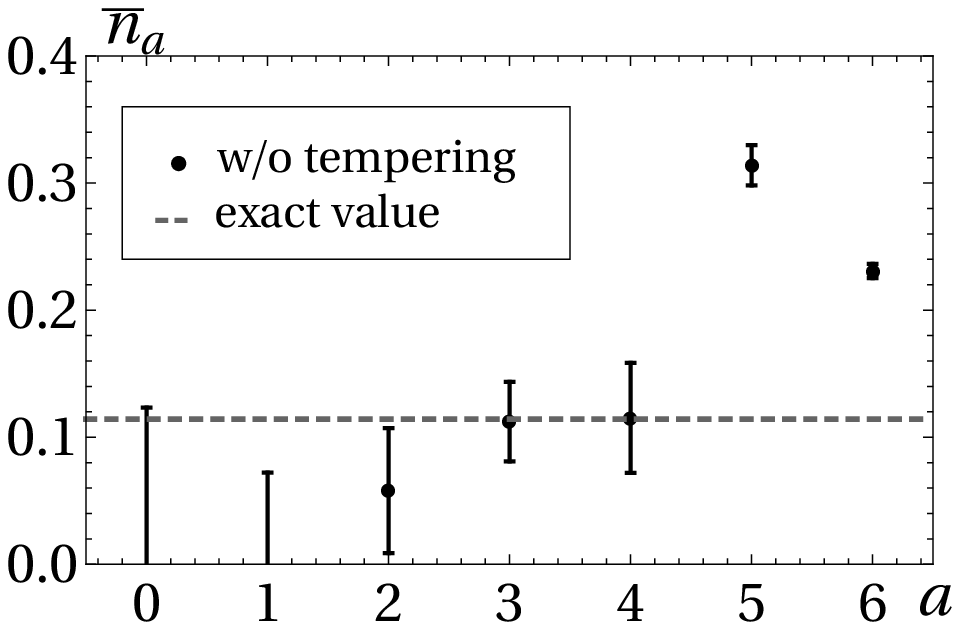} 
  \caption{
    \label{fig:met_results_noswap}
    The results for Metropolis without swap. 
    (Left) The sign averages. 
    (Right) The estimates $\bar{n}_a$. 
  }
\end{figure}

We also plot the sign averages and the expectation values 
obtained {\it without} swap 
in Figs.~\ref{fig:hmc_results_noswap} and \ref{fig:met_results_noswap}. 
\noindent
\noindent
For large $t$, we observe significant deviations 
from the exact value but small error margins, 
reflecting the presence of the ergodicity problem \cite{Fukuma:2019wbv}. 

\subsection{Autocorrelations}
\label{sec:autocorrelations}

In this subsection, 
we evaluate the autocorrelations for both HMC and Metropolis, 
and compare their efficiencies. 
For the evaluation, 
we estimate the normalized autocorrelation function by 
(see, e.g., \cite{Priestley:1981,Madras:1988ei})
\begin{align}
  \rho(m) = \frac{C(m)}{C(0)}. 
  \label{eq:normalized_autocorrelation_function}
\end{align}
Here, $C(m)$ is given by 
\begin{align}
  C(m) \equiv
  \frac{1}{{N_{\rm conf}}-m} \sum_{k=1}^{{N_{\rm conf}}-m}
  [
  f(z^{(k)}_a) - \overline{f(z_a)}
  ]
  [
  f(z^{(k+m)}_a) - \overline{f(z_a)}
  ],
\end{align}
where $\overline{f(z_a)}$ is a sample average 
for the subsample at replica $a$, 
$\overline{f(z_a)}\equiv(1/{N_{\rm conf}})\sum_{k=1}^{{N_{\rm conf}}}f(z^{(k)}_a)$ 
(see subsection~\ref{sec:TLTM}). 
Then, we estimate the integrated autocorrelation time 
by the following formula, 
which is valid for 
$\tau_\mathrm{int} \ll k_{\max} \ll {N_{\rm conf}}$ \cite{Madras:1988ei}: 
\begin{align}
  \tau_\mathrm{int} = 1 + 2\sum_{k=1}^{k_{\max}}\rho(k). 
  \label{eq:tauint}
\end{align}
When we plot the right-hand side of eq.~(\ref{eq:tauint}) 
as a function of $k_{\max}$, 
we expect to observe a plateau. 
In the following, 
we choose the plateau region manually, 
and define the estimate of $\tau_{\rm int}$ 
to be the value of $\tau_{\rm int}$ at the least $k_{\max}$ in the region. 
Note that $\rho(m)$ and $\tau_\mathrm{int}$ depend
on the choice of operators and replicas. 
We apply the above formulas to the operators 
$f(z) = \mathrm{Re}\,[ e^{i\theta(z)}  n(z) ]$ 
and $f(z)= \mathrm{Re}\,[ e^{i\theta(z)} ]=\cos\theta(z)$ at replica $a=A$. 

\begin{figure}[ht]
  \centering
  \includegraphics[width=60mm]{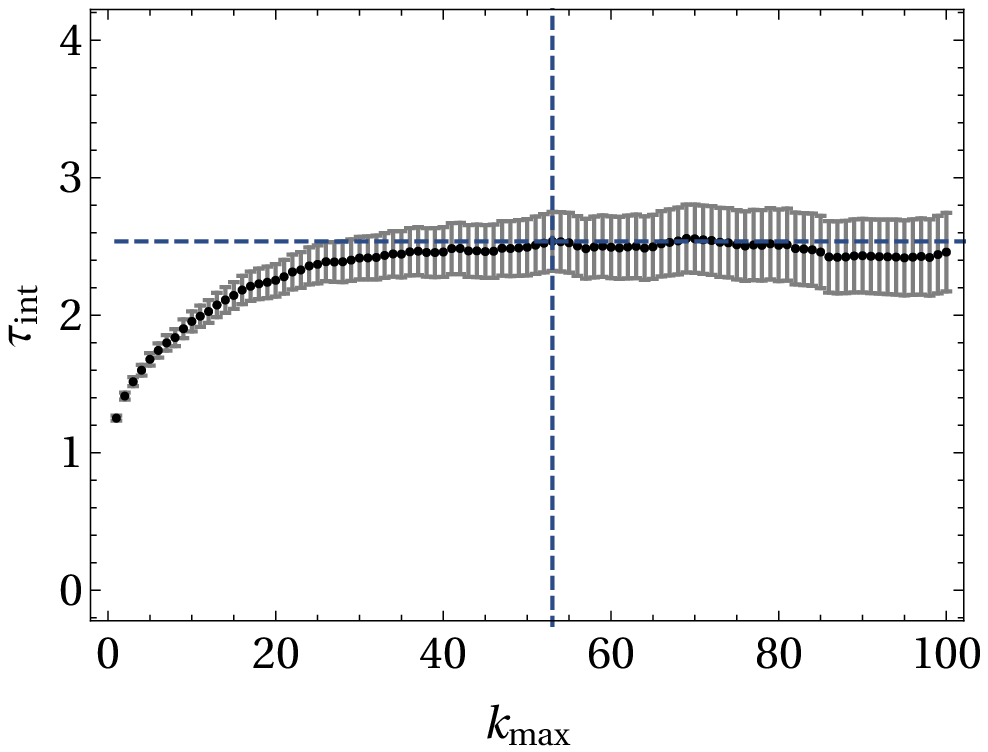}%
  \hspace{1cm}
  \includegraphics[width=60mm]{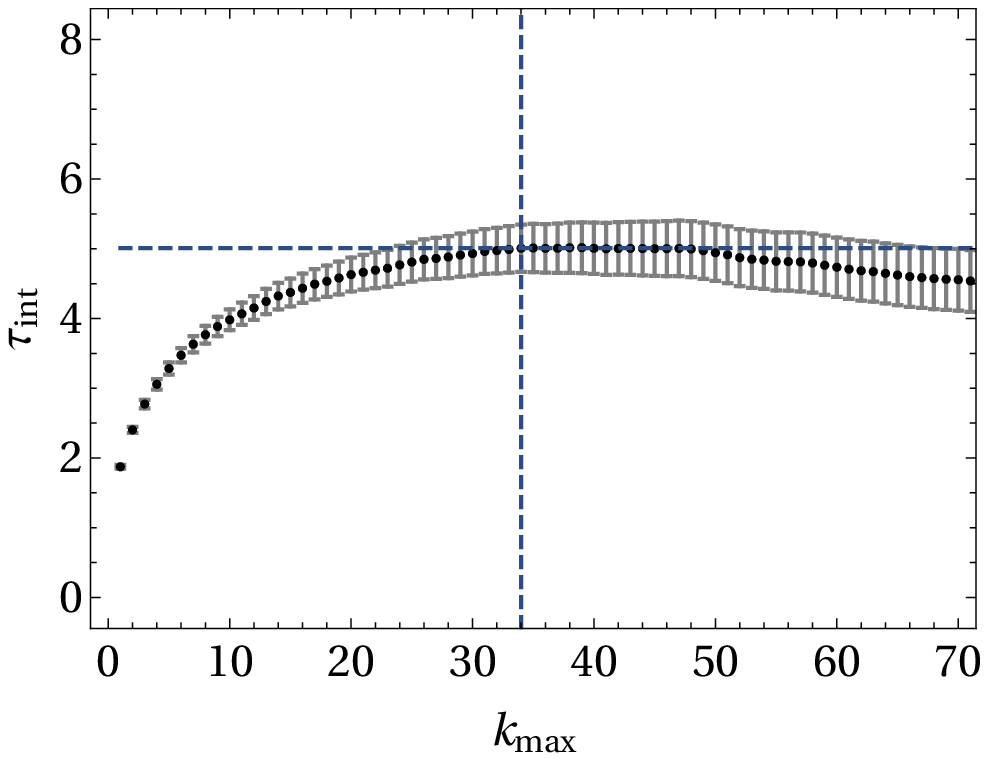} 
  \caption{
    \label{fig:tauint_numer}
    The estimates of $\tau_\mathrm{int}$ 
    for $\mathrm{Re}\,[ e^{i\theta(z)} n(z)]$ (with swap). 
    (Left) HMC. 
    (Right) Metropolis. 
    The horizontal dashed line indicates the value of $\tau_\mathrm{int}$, 
    and the vertical dashed line indicates the value of $k_{\max}$, 
  }
\end{figure}

\begin{figure}[ht]
  \centering
  \includegraphics[width=60mm]{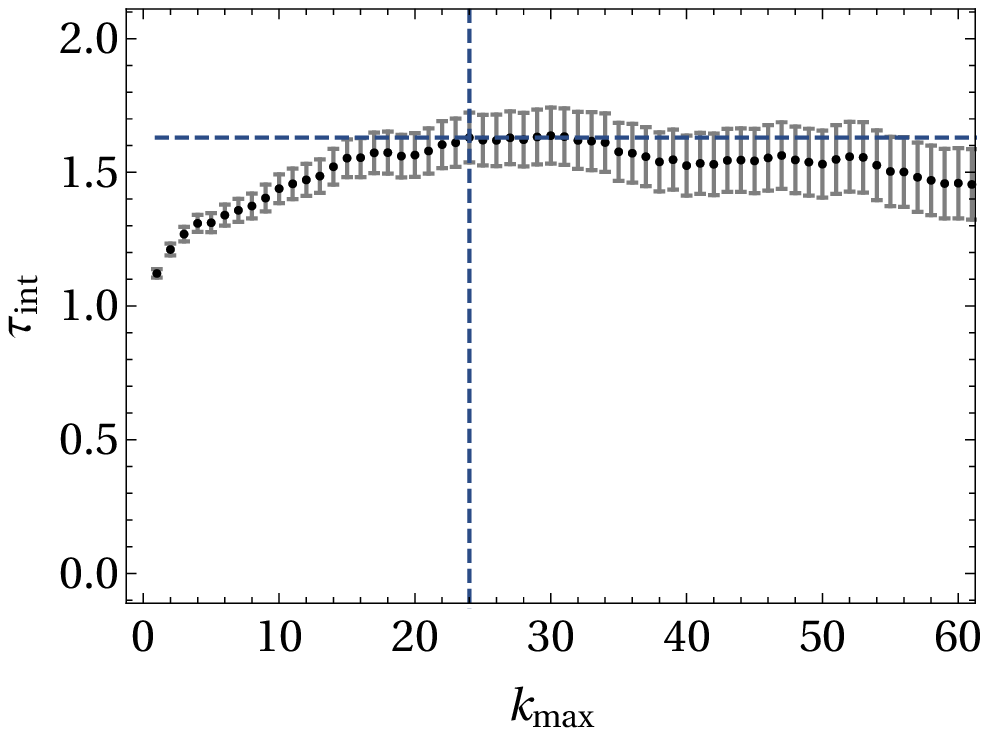}%
  \hspace{1cm}
  \includegraphics[width=60mm]{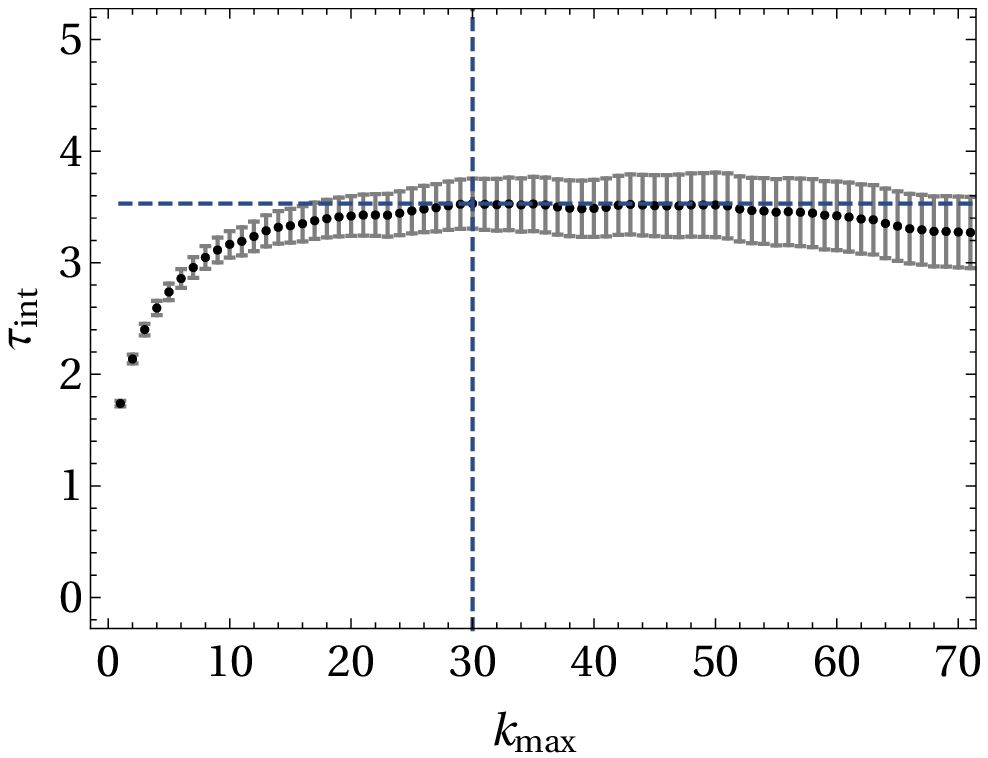} 
  \caption{
    \label{fig:tauint_denom}
    The estimates of $\tau_\mathrm{int}$ 
    for $\cos\theta(z)$ (with swap). 
    (Left) HMC. 
    (Right) Metropolis. 
  }
\end{figure}

We first investigate the autocorrelation 
for $f(z) = \mathrm{Re}\,[ e^{i\theta(z)}  n(z) ]$. 
Figure \ref{fig:tauint_numer} shows $\tau_\mathrm{int}$ for various $k_{\max}$, 
from which we identify the plateau 
and estimate $\tau_{\rm int}$ as given 
in Table \ref{table:autocorrelation_results}. 
\noindent
\begin{table}[ht]
  \centering
  \begin{tabular}{|c|c|c|c|}
    \hline
    $f(z)$ & algorithm & $\tau_\mathrm{int}$ & $k_{\max}$  \\
    \hline
    $\mathrm{Re}\,[ e^{i\theta(z)} n(z) ]$
           & HMC & $2.54 \pm 0.21$ & 53 \\
           & Metropolis & $5.01 \pm 0.34$ & 34 \\
    \hline
    $\cos\theta(z)$
           & HMC & $1.630 \pm 0.093$ & 24 \\
           & Metropolis & $3.53 \pm 0.23$ & 30 \\
    \hline
  \end{tabular}
  \caption{The estimates of $\tau_\mathrm{int}$ (with swap).}
  \label{table:autocorrelation_results}
\end{table}\noindent
We see that $\tau_{\rm int}$ for HMC is 
about $50\%$ of that for Metropolis 
with respect to this operator.
A similar analysis is carried out for $f(z) = \cos\theta(z)$ 
(see Fig.~\ref{fig:tauint_denom}). 
\noindent
The estimates of $\tau_\mathrm{int}$ are also given  
in Table \ref{table:autocorrelation_results}. 
We see that $\tau_{\rm int}$ for HMC is 
also about $50\%$ of that for Metropolis 
with respect to this operator. 

As a comparison of the actual efficiency between the two algorithms, 
we comment that the elapsed time to obtain a single configuration 
is in average $7.8~\mathrm{sec}$
for HMC
and $15~\mathrm{sec}$ for Metropolis.%
\footnote{
  The estimations were made 
  by using Intel Xeon E5-2667 v4 running at 3.2 GHz 
  with seven threads for each algorithm. 
\label{fn:Xeon}
} 
Therefore, 
the actual computational cost to obtain one independent configuration with HMC
is about $ 30 \% $ of that with Metropolis. 
We expect that the benefits in computational cost 
become more significant as the degrees of freedom increase. 

\begin{figure}[ht]
  \centering
  \includegraphics[width=60mm]{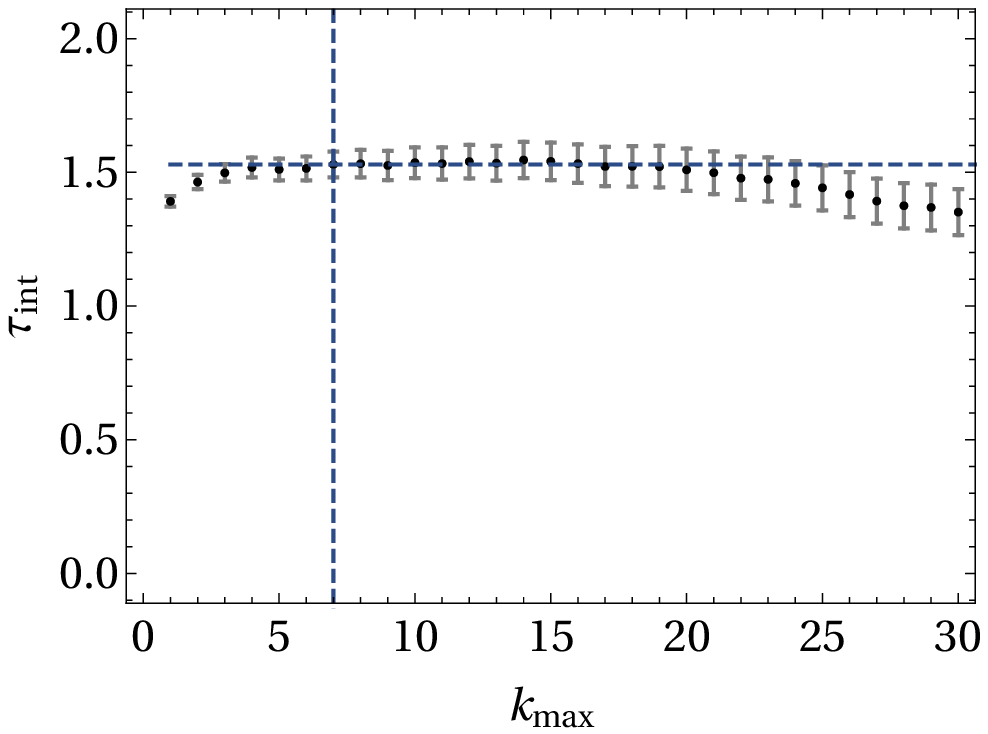}%
  \hspace{1cm}
  \includegraphics[width=60mm]{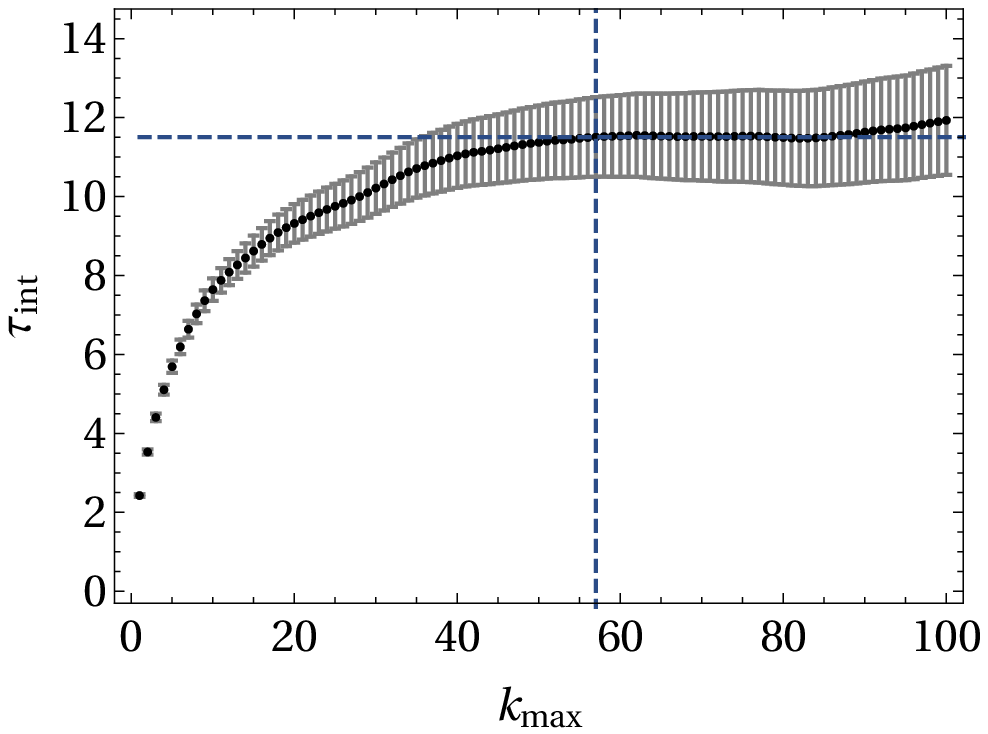} 
  \caption{
    \label{fig:tauint_numer_noswap}
    The estimates of $\tau_\mathrm{int}$ 
    for $\mathrm{Re}\,[ e^{i\theta(z)} n(z)]$ (without swap). 
    (Left) HMC. 
    (Right) Metropolis. 
  }
\end{figure}

\begin{figure}[ht]
  \centering
  \includegraphics[width=60mm]{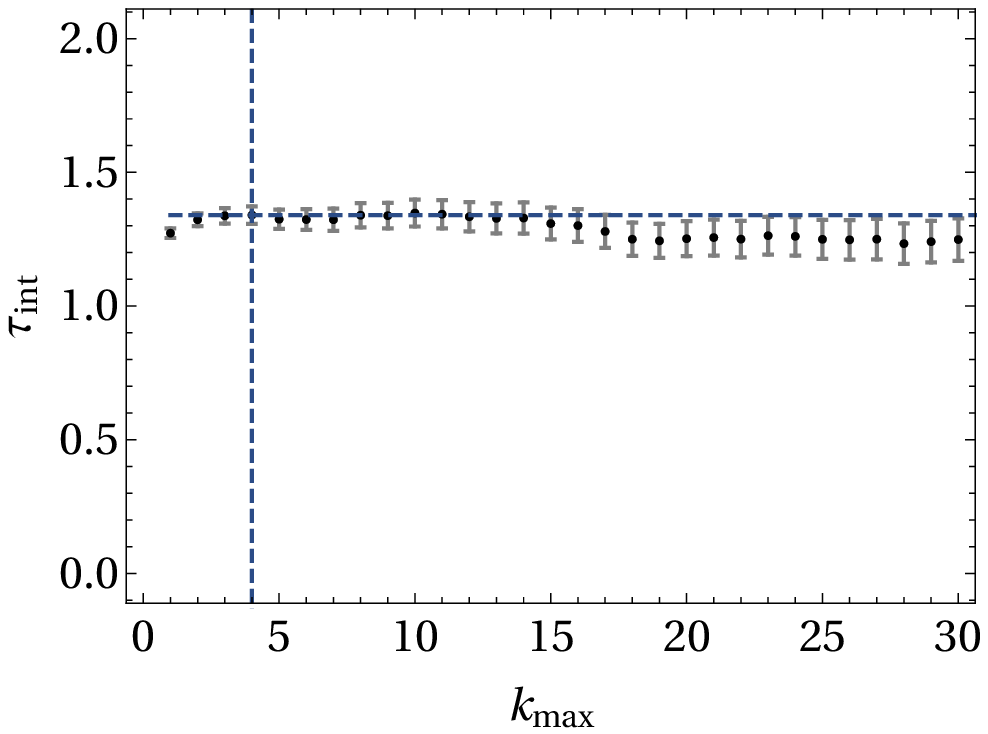}%
  \hspace{1cm}
  \includegraphics[width=60mm]{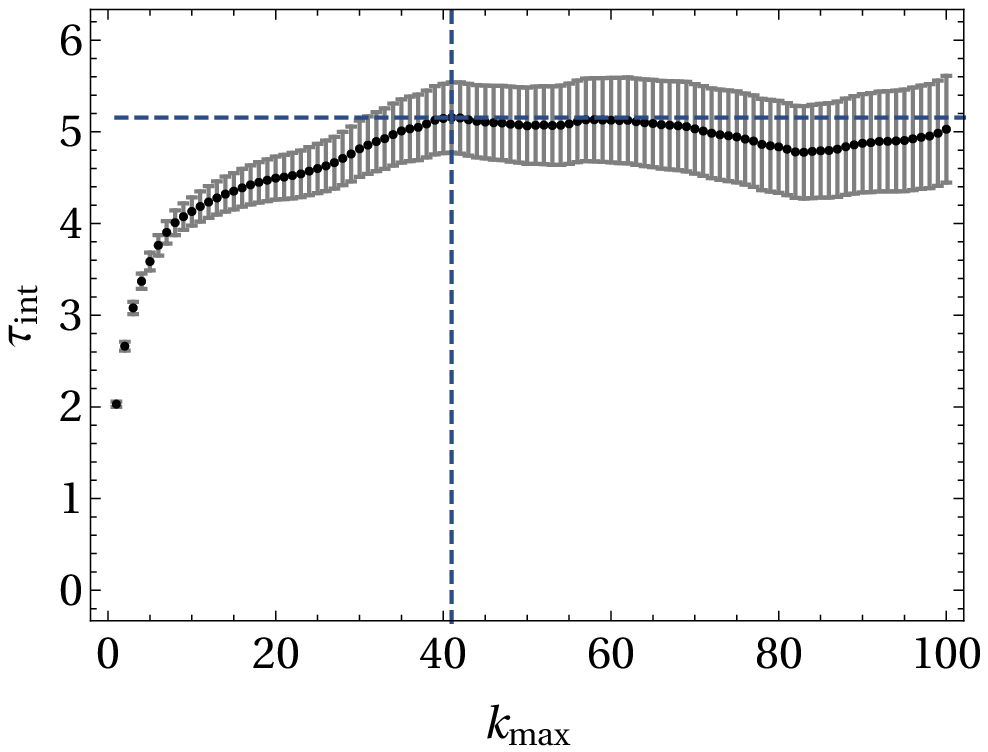} 
  \caption{
    \label{fig:tauint_denom_noswap}
    The estimates of $\tau_\mathrm{int}$ for $\cos\theta(z)$ (without swap). 
    (Left) HMC. 
    (Right) Metropolis. 
  }
\end{figure}

We here comment that the difference of $\tau_{\rm int}$
between HMC and Metropolis 
becomes more significant for simulations {\it without} swap. 
In Table \ref{table:autocorrelation_results_noswap} 
we summarize the results obtained without swap. 
\begin{table}[ht]
  \centering
  \begin{tabular}{|c|c|c|c|}
    \hline
    $f(z)$ & algorithm & $\tau_\mathrm{int}$ & $k_{\max}$  \\
    \hline
    $\mathrm{Re}\,[ e^{i\theta(z)} n(z)]$
           & HMC & $1.529 \pm 0.048$ & 7 \\
           & Metropolis & $11.5 \pm 1.0$ & 57 \\
    \hline
    $\cos\theta(z)$
           & HMC & $1.340 \pm 0.033$ & 4 \\
           & Metropolis & $5.16 \pm 0.38$ & 41 \\
    \hline
  \end{tabular}
  \caption{The estimates of $\tau_\mathrm{int}$ (without swap).}
  \label{table:autocorrelation_results_noswap}
\end{table}\noindent
The corresponding plots
for $f(z) = \mathrm{Re}\,[ e^{i\theta(z)}  n(z) ]$ 
are shown in 
Fig.~\ref{fig:tauint_numer_noswap}, 
and those for $f(z) = \cos\theta(z)$ 
in Fig.~\ref{fig:tauint_denom_noswap}. 
\noindent
\noindent
We see that $\tau_{\rm int}$ for HMC 
is $ 13-26 \% $ of those for Metropolis . 
The elapsed time to obtain a single configuration 
is in average 
$4.1~\mathrm{sec}$ for HMC  
and $12~\mathrm{sec}$ for Metropolis.%
\footnote{
  The estimations were made with the same environment as footnote~\ref{fn:Xeon}.
} 
Thus the actual computational cost to obtain one independent configuration 
with HMC
is less than $10 \%$ of that with Metropolis 
for the calculations without swap.

\section{Conclusion and outlook}
\label{sec:conclusion}

In this paper, we implemented the HMC algorithm on the TLTM, 
aiming to apply our algorithm 
to systems including fermions 
with large degrees of freedom. 
We observed that the actual computational cost to obtain 
an independent configuration 
becomes about 30\% of that for the Metropolis algorithm 
even for such small degrees of freedom ($N=20$). 

We expect that the above improvement makes the TLTM 
more effective in solving the sign problems listed in Introduction, 
especially when performed on a large-scale computer. 
In parallel with the application of the algorithm to those problems 
(and also to some simplified model 
such as chiral random matrix models \cite{Stephanov:1996ki,Halasz:1998qr}),  
it should be important 
to further develop the algorithm itself. 
In particular, the following three issues should be addressed: 

(1) It is desirable to have 
a systematic method to estimate numerical errors 
introduced in integrating the antiholomorphic gradient flow 
and in solving Newton's method iteratively 
(Step 1 in subsection~\ref{sec:MD_Sigma_t}). 

(2) We should investigate the scaling of computational cost 
as the degrees of freedom are increased. 
A simple estimate of the total cost of our algorithm is $O(N^{3-4})$. 
$O(N^3)$ comes from the calculation of the Jacobian, 
and $O(N^{0-1})$ comes from the need to increase the number of replicas 
to keep the acceptance rates at swapping to significant values. 
It should be crucial to investigate 
if the above scaling is actually realized in large-scale calculations, 
because it then means that we can obtain correct results 
with a computational cost of a power of $N$ 
(not an exponential of $N$). 

(3) It should be helpful to have a systematic understanding 
of the global sign problem 
(i.e.\ cancellations of phases among different thimbles) 
and the residual sign problem 
(i.e.\ contributions from the phase factor $e^{i\varphi(z)}$) 
for systems with large degrees of freedom, 
because they can be a cause of 
a significant increase of computational cost. 

A study along these lines is now in progress and will be reported elsewhere.

\section*{Acknowledgments}
The authors thank Andrei 
Alexandru, Gerald Dunne, Hitotsugu Fujii, Yoshimasa Hidaka, 
Ken-Ichi Ishikawa, Issaku Kanamori, Masakiyo Kitazawa, 
Yoshifumi Nakamura, Yusuke Namekawa, Jun Nishimura, 
Akira Ohnishi, Yusuke Taniguchi and Shoichiro Tsusui
for useful discussions, 
and especially Yoshio Kikukawa 
for sharing his insights on the RATTLE process 
in Lefschetz thimble methods. 
This work was partially supported by JSPS KAKENHI 
(Grant Numbers 16K05321, 18J22698 and 17J08709) 
and by SPIRITS 2019 of Kyoto University (PI: M.F.).

\appendix

\section{More on the treatment of zeros}
\label{sec:more_on_zeros} 

In subsection \ref{sec:MD_Sigma_t}, 
we constructed a molecular dynamics on $\Sigma_t$, 
where a point $z=z_t(x)$ moves to another point $z'=z_t(x+u)$, 
and the increment $u=(u^\alpha)\in\bbR^N$ is obtained 
by solving \eqref{ul_eq} 
with respect to $w=(u^\alpha,\lambda^\alpha)^T\in\bbR^{2N}$. 

\begin{figure}[ht]
  \centering
  \includegraphics[width=52mm]{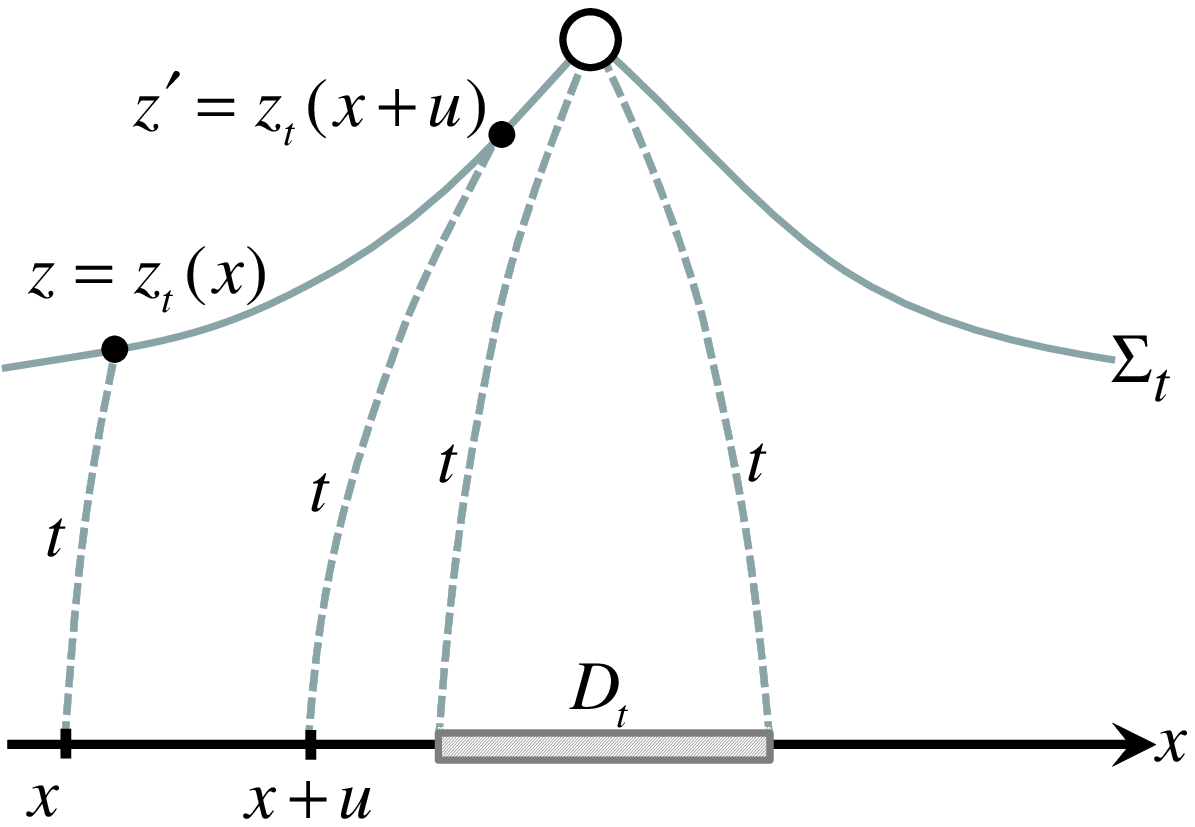}%
  \hspace{10mm}
  \includegraphics[width=52mm]{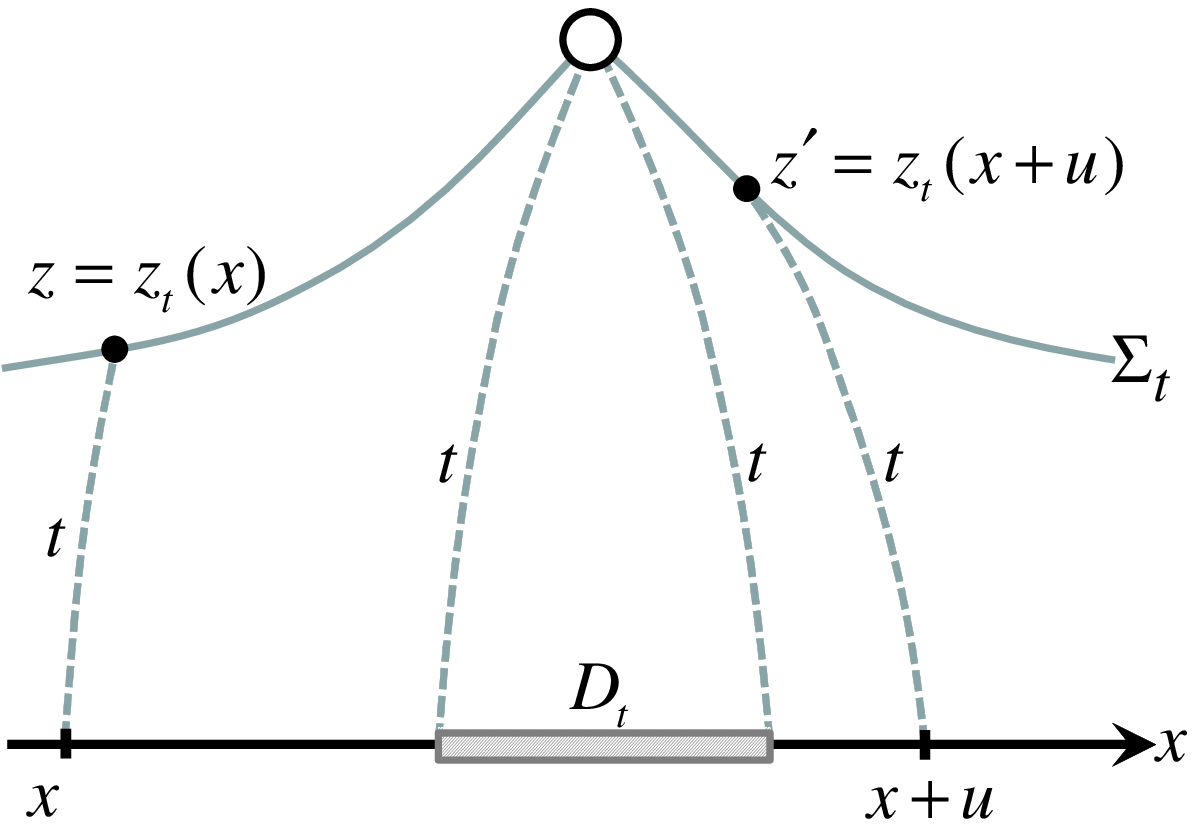} 
  \caption{
    \label{fig:Newton}
    Possible solutions to \eqref{ul_eq}. 
    $D_t$ is the domain consisting of the points 
    that flow to a zero (denoted by a circle) with flow times less than $t$. 
    $x+u$ is on the same side as $x$ (left), 
    or in a region beyond $D_t$ (right). 
  }
\end{figure}

\begin{figure}[ht]
  \centering
  \includegraphics[width=52mm]{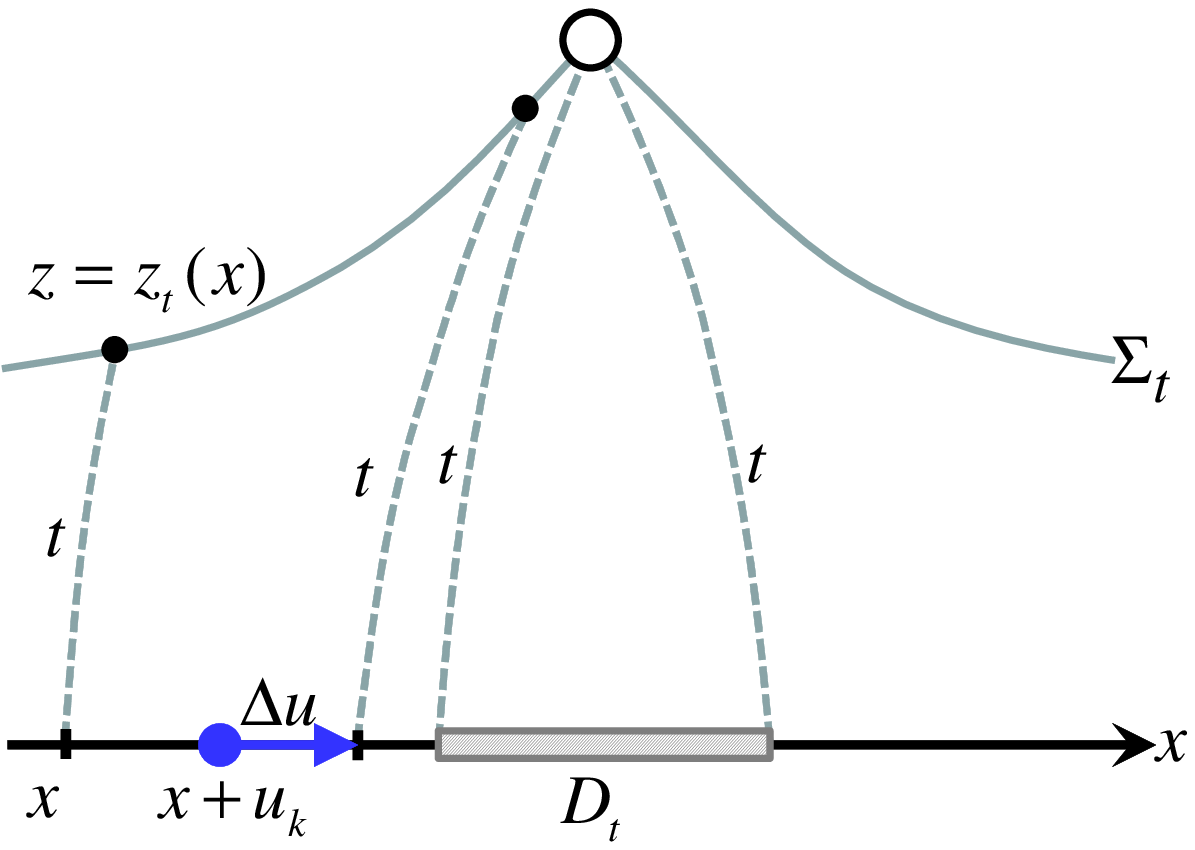}%
  \hspace{3mm}
  \includegraphics[width=52mm]{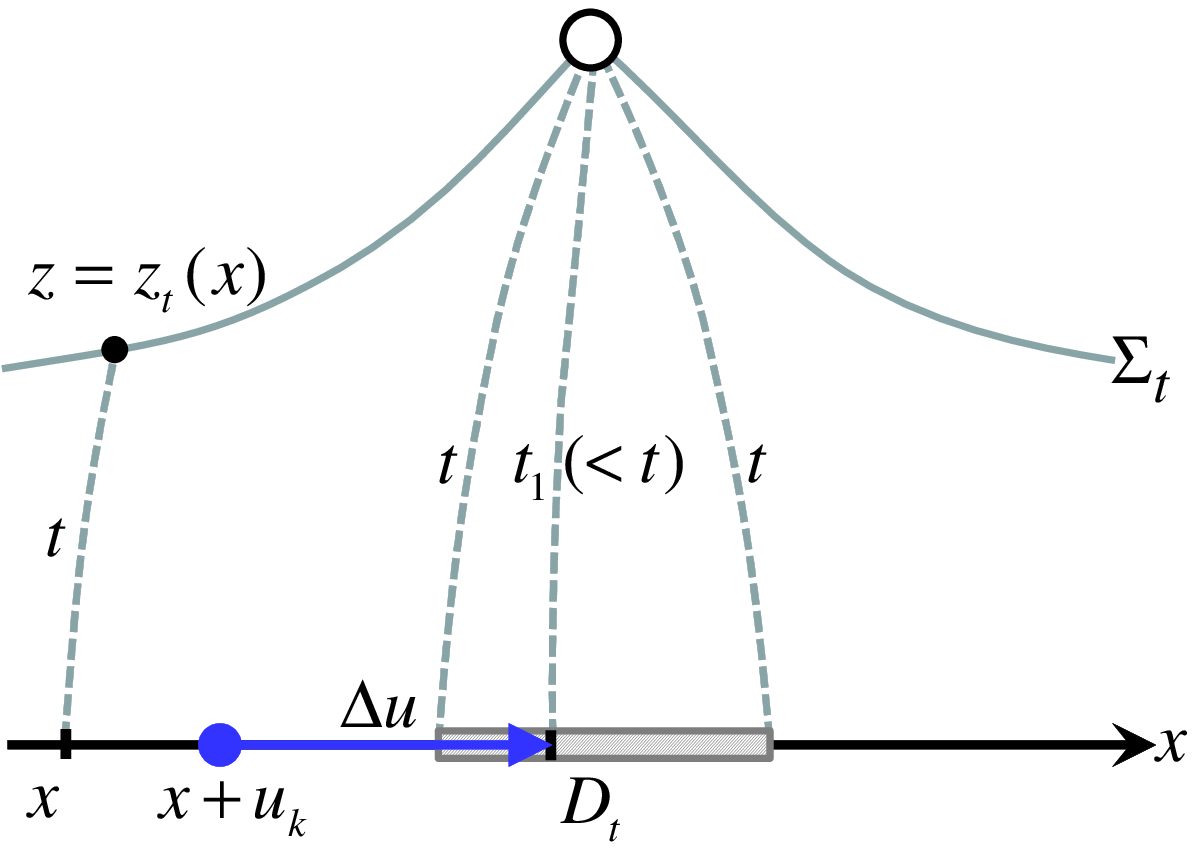}%
  \hspace{3mm}
  \includegraphics[width=52mm]{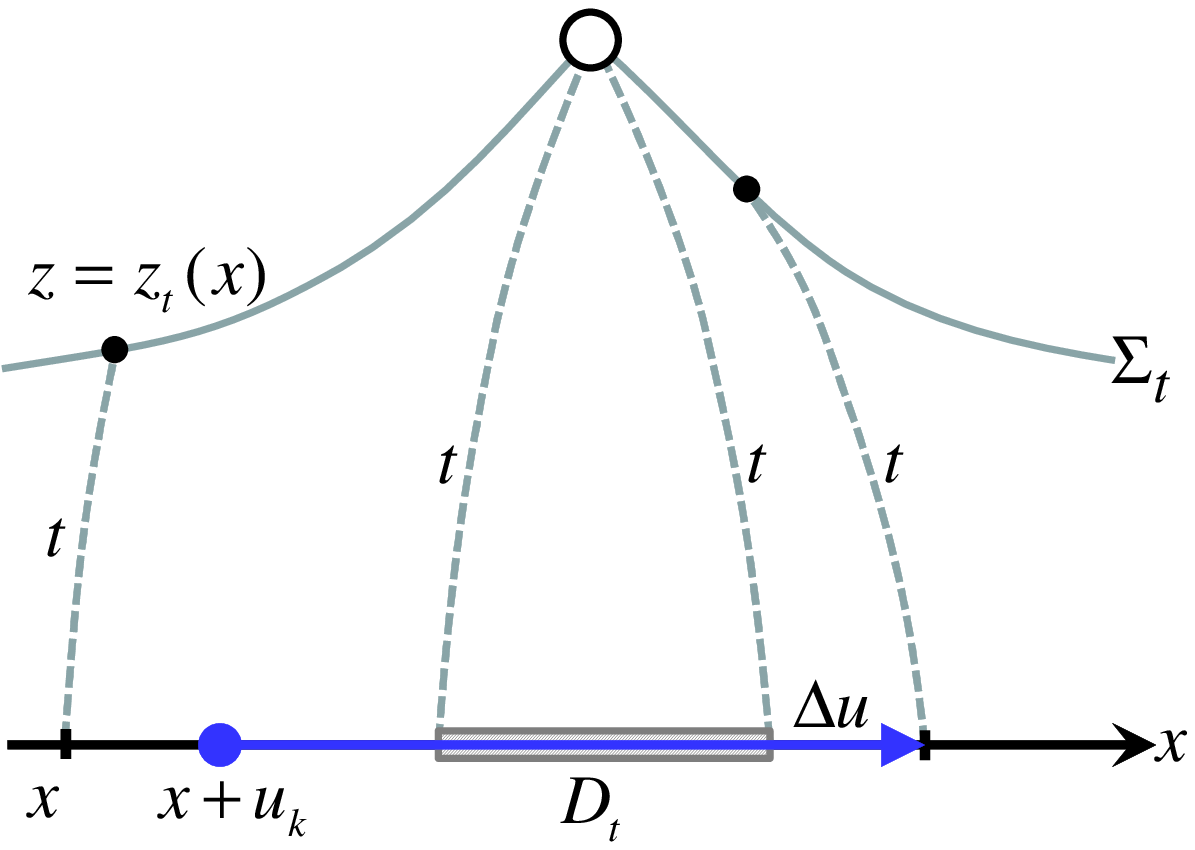} 
  \caption{
    \label{fig:iteration}
    An iterative step in Newton's method giving $u_{k+1}=u_k+\Delta u$. 
    $x+u_{k+1}$ can be on the same side as $x$ (left), 
    in the domain $D_t$ (middle), 
    or in a region beyond $D_t$ (right). 
  }
\end{figure}

Among possible solutions for $w=(u,\lambda)^T$, 
we look for the solution 
that gives $z'=z_t(x+u)$ closest to $z$, 
which will ensure the molecular-dynamics step 
to satisfy the reversibility. 
Since $u=O(\Delta s)$ and $\lambda=O(\Delta s^2)$ 
($\Delta s$: step size of molecular dynamics), 
we would expect that such solution can be found uniquely 
by setting $\Delta s$ to a sufficiently small value 
and by iteratively solving \eqref{ul_eq} with Newton's method, 
$w_k\to w_{k+1}=w_k+\Delta w$ [see \eqref{recursion2}], 
with the initial guess $w_0=0$. 
However, special attention needs to be paid 
when $z=z_t(x)$ is near a zero of the weight $e^{-S(z)}$, 
because this means that the initial configuration $x$ 
is close to a domain $D_t$ 
that consists of the points flowing to the zero 
with flow times less than $t$ (see Fig.~\ref{fig:Newton}). 
Then, there can be a solution that gives $x+u$ in a region beyond $D_t$ 
(see the right panel of Fig.~\ref{fig:Newton}). 
However, this solution is not what we are looking for 
because this can give $u$ much larger than $O(\Delta s)$. 
\noindent
Such a case can be identified 
if we observe that 
the sequence $x+u_k+\Delta u$ enters the domain $D_t$ 
while keeping the increment $\Delta w$ (and thus $\Delta u$) 
to small values 
so that the sequence does not leap over $D_t$ 
(see Fig.~\ref{fig:iteration}). 
\noindent
Note that there can also be a case 
where a solution is not found anywhere. 

The above consideration leads us to the following algorithm:

\begin{description}

\item[Step 1.]
We start from the initial guess $w_0=0$, 
and at every step $w_k\to w_{k+1}=w_k+\Delta w$, 
we rescale the obtained $\Delta w$ 
as $\Delta w \rightarrow 0.15 \times \Delta w$ 
unless $|\Delta w|/\sqrt{2N} < 0.5\times\Delta s/|\det J_t(x)|^{1/N}$ 
and $|f(w_k+\Delta w)| < |f(w_k)|$.%
\footnote{
  The requirement with the first inequality is that 
  the ``typical magnitude'' of $w^I$ $(I=1,\ldots,2N)$
  (which we estimate to be $|w|/\sqrt{2N}$) 
  be less than $0.5\,\Delta s$ 
  even after it is stretched under the flow 
  (by a factor which we estimate to be $|\det J_t(x)|^{1/N}$). 
  The second inequality prohibits the sequence to go away from a nearby solution.
} 

\item[Step 2.]
We continue the iterative steps 
until we reach one of the following three cases:\\
~~(A)
The sequence converges to a solution.\\
~~(B)
The sequence enters the domain $D_t$.\\
~~(C)
The sequence will not converge to any solution.\\
As a criterion for (A), 
we use the conditions 
$|\Delta w|/\sqrt{2N}<10^{-8}$ and $|f(w_k)|/\sqrt{2N}<10^{-5}$. 
Case (B) is identified 
if $x+u_k+\Delta u$, at some step $k$, flows to a zero 
with a flow time less than $t$
even after the rescaling 
$\Delta w \rightarrow 0.15^3 \times \Delta w$. 
Case (C) is identified 
if the number of iterations exceeds 50. 
As a comparison, 
we find that the process terminates with $3-7$ iterations for case (A). 

\item[Step 3.]
For case (A), we set $z'\equiv z_t(x+u)\in\Sigma_t$ 
and calculate $\pi'\in T^\ast\Sigma_t$ 
to get $(z',\pi')=\Phi_{\Delta s}(z,\pi)$. 
For cases (B) and (C), 
we stop the iteration 
and set $(z',\pi')\equiv (z,-\pi)\,(= \Psi(z,\pi))$, 
i.e., we use the momentum flip $\Psi$ instead of $\Phi_{\Delta s}$. 

\end{description}

\baselineskip=0.9\normalbaselineskip



\end{document}